\documentclass[twocolumn]{aastex631}

\newcommand\fakesection[1]{%
  \refstepcounter{section}%
  \addcontentsline{toc}{section}{\protect\numberline{\thesection}#1}%
  \sectionmark{#1}}

\begin{document}

\title{Transmission Spectroscopy of the Habitable Zone Exoplanet LHS~1140\,b with JWST/NIRISS}

\correspondingauthor{Charles Cadieux}
\email{charles.cadieux.1@umontreal.ca}

\author[0000-0001-9291-5555]{Charles Cadieux}
\affiliation{Institut Trottier de recherche sur les exoplanètes, Université de Montréal, 1375 Ave Thérèse-Lavoie-Roux, Montréal, QC, H2V 0B3, Canada}

\author[0000-0001-5485-4675]{Ren\'e Doyon}
\affiliation{Institut Trottier de recherche sur les exoplanètes, Université de Montréal, 1375 Ave Thérèse-Lavoie-Roux, Montréal, QC, H2V 0B3, Canada}
\affiliation{Observatoire du Mont-M\'egantic, Universit\'e de Montr\'eal, Montr\'eal H3C 3J7, Canada}

\author[0000-0003-4816-3469]{Ryan J. MacDonald}
\affiliation{Department of Astronomy, University of Michigan, 1085 S. University Ave., Ann Arbor, MI 48109, USA}
\affiliation{NHFP Sagan Fellow}

\author[0000-0003-2260-9856]{Martin Turbet}
\affiliation{Laboratoire de M\'et\'eorologie Dynamique/IPSL, CNRS, Sorbonne Universit\'e, Ecole Normale Sup\'erieure, PSL Research University, Ecole Polytechnique, 75005 Paris, France}
\affiliation{Laboratoire d'astrophysique de Bordeaux, Univ. Bordeaux, CNRS, B18N, allée Geoffroy Saint-Hilaire, 33615 Pessac, France}

\author[0000-0003-3506-5667]{\'Etienne Artigau}
\affiliation{Institut Trottier de recherche sur les exoplanètes, Université de Montréal, 1375 Ave Thérèse-Lavoie-Roux, Montréal, QC, H2V 0B3, Canada}
\affiliation{Observatoire du Mont-M\'egantic, Universit\'e de Montr\'eal, Montr\'eal H3C 3J7, Canada}

\author[0000-0003-4676-0622]{Olivia Lim}
\affiliation{Institut Trottier de recherche sur les exoplanètes, Université de Montréal, 1375 Ave Thérèse-Lavoie-Roux, Montréal, QC, H2V 0B3, Canada}

\author[0000-0002-3328-1203]{Michael Radica}
\affiliation{Institut Trottier de recherche sur les exoplanètes, Université de Montréal, 1375 Ave Thérèse-Lavoie-Roux, Montréal, QC, H2V 0B3, Canada}

\author[0000-0002-5967-9631]{Thomas J. Fauchez}
\affiliation{NASA Goddard Space Flight Center
8800 Greenbelt Road Greenbelt, MD 20771, USA}
\affiliation{Integrated Space Science and Technology Institute, Department of Physics, American University, Washington DC}
\affiliation{NASA GSFC Sellers Exoplanet Environments Collaboration}

\author[0000-0001-6758-7924]{Salma Salhi}
\affiliation{Institut Trottier de recherche sur les exoplanètes, Université de Montréal, 1375 Ave Thérèse-Lavoie-Roux, Montréal, QC, H2V 0B3, Canada}

\author[0000-0003-4987-6591]{Lisa Dang}
\affiliation{Institut Trottier de recherche sur les exoplanètes, Université de Montréal, 1375 Ave Thérèse-Lavoie-Roux, Montréal, QC, H2V 0B3, Canada}

\author[0000-0003-0475-9375]{Lo\"ic Albert}
\affiliation{Institut Trottier de recherche sur les exoplanètes, Université de Montréal, 1375 Ave Thérèse-Lavoie-Roux, Montréal, QC, H2V 0B3, Canada}

\author[0000-0002-2195-735X]{Louis-Philippe Coulombe}
\affiliation{Institut Trottier de recherche sur les exoplanètes, Université de Montréal, 1375 Ave Thérèse-Lavoie-Roux, Montréal, QC, H2V 0B3, Canada}

\author[0000-0001-6129-5699]{Nicolas B. Cowan}
\affiliation{Department of Earth \& Planetary Sciences, McGill University, 3450 rue University, Montréal, QC H3A 0E8, Canada}
\affiliation{Department of Physics and Trottier Space Institute, McGill University, Montréal, Québec, H3A 2T8, Canada}

\author[0000-0002-6780-4252]{David Lafreni\`ere}
\affiliation{Institut Trottier de recherche sur les exoplanètes, Université de Montréal, 1375 Ave Thérèse-Lavoie-Roux, Montréal, QC, H2V 0B3, Canada}

\author[0009-0005-6135-6769]{Alexandrine L'Heureux}
\affiliation{Institut Trottier de recherche sur les exoplanètes, Université de Montréal, 1375 Ave Thérèse-Lavoie-Roux, Montréal, QC, H2V 0B3, Canada}

\author[0000-0002-2875-917X]{Caroline Piaulet-Ghorayeb}
\affiliation{Institut Trottier de recherche sur les exoplanètes, Université de Montréal, 1375 Ave Thérèse-Lavoie-Roux, Montréal, QC, H2V 0B3, Canada}

\author[0000-0001-5578-1498]{Bj\"orn Benneke}
\affiliation{Institut Trottier de recherche sur les exoplanètes, Université de Montréal, 1375 Ave Thérèse-Lavoie-Roux, Montréal, QC, H2V 0B3, Canada}

\author[0000-0001-5383-9393]{Ryan Cloutier}
\affiliation{Department of Physics \& Astronomy, McMaster University, 1280 Main St W, Hamilton, ON, L8S 4L8, Canada}

\author[0000-0003-0977-6545]{Benjamin Charnay}
\affiliation{LESIA, Observatoire de Paris, Université PSL, CNRS, Sorbonne Université, Université Paris-Cité, 5 place Jules Janssen, 92195 Meudon, France}

\author[0000-0003-4166-4121]{Neil J. Cook}
\affiliation{Institut Trottier de recherche sur les exoplanètes, Université de Montréal, 1375 Ave Thérèse-Lavoie-Roux, Montréal, QC, H2V 0B3, Canada}

\author[0000-0002-5428-0453]{Marylou Fournier-Tondreau}
\affiliation{Atmospheric Oceanic and Planetary Physics, Department of Physics, University of Oxford, OX1 3PU, UK}

\author[0000-0002-9479-2744]{Mykhaylo Plotnykov}
\affiliation{Department of Physics, University of Toronto, Toronto, ON M5S 3H4, Canada}

\author[0000-0003-3993-4030]{Diana Valencia}
\affiliation{Department of Physical \& Environmental Sciences, University of Toronto at Scarborough, Toronto, ON M1C 1A4, Canada}
\affiliation{David A. Dunlap Dept.\ of Astronomy \& Astrophysics, University of Toronto, 50 St. George Street, Toronto, Ontario, M5S 3H4, Canada}

\begin{abstract}
LHS~1140\,b is the second-closest temperate transiting planet to the Earth with an equilibrium temperature low enough to support surface liquid water. At 1.730$\pm$0.025\,R$_\oplus$, LHS~1140\,b falls within the radius valley separating H$_2$-rich mini-Neptunes from rocky super-Earths. Recent mass and radius revisions indicate a bulk density significantly lower than expected for an Earth-like rocky interior, suggesting that LHS~1140\,b could either be a mini-Neptune with a small envelope of hydrogen ($\sim$0.1\% by mass) or a water world (9--19\% water by mass). Atmospheric characterization through transmission spectroscopy can readily discern between these two scenarios. Here, we present two JWST/NIRISS transit observations of LHS~1140\,b, one of which captures a serendipitous transit of LHS~1140\,c. The combined transmission spectrum of LHS~1140\,b shows a telltale spectral signature of unocculted faculae (5.8\,$\sigma$), covering $\sim$20\% of the visible stellar surface. Besides faculae, our spectral retrieval analysis reveals tentative evidence of residual spectral features, best-fit by Rayleigh scattering from an N$_2$-dominated atmosphere (2.3\,$\sigma$), irrespective of the consideration of atmospheric hazes. We also show through Global Climate Models (GCM) that H$_2$-rich atmospheres of various compositions (100$\times$, 300$\times$, 1000$\times$solar metallicity) are ruled out to $>$10\,$\sigma$. The GCM calculations predict that water clouds form below the transit photosphere, limiting their impact on transmission data. Our observations suggest that LHS~1140\,b is either airless or, more likely, surrounded by an atmosphere with a high mean molecular weight. Our tentative evidence of an N$_2$-rich atmosphere provides strong motivation for future transmission spectroscopy observations of LHS~1140\,b.
\end{abstract}

\section{Introduction} \label{sec:intro}

Whether temperate rocky planets orbiting low-mass stars have an atmosphere is arguably one of the most important scientific questions of the James Webb Space Telescope (JWST) mission \citep{Gardner_2023}, one that can only be answered if a significant fraction of its lifetime is dedicated to this endeavour \citep{Cowan_2015, dewit2023, Doyon_2024}.  Answering this question is an essential first step in assessing the habitability of nearby worlds and identifying the best targets for biosignature searches. The first 18 months of JWST observations have highlighted both its power and versatility for studying the atmospheres of small exoplanets. While the first atmospheric reconnaissance of TRAPPIST-1\,b \citep{Greene_2023} and TRAPPIST-1\,c \citep{Zieba_2023} through eclipse photometry suggests that those two planets may be airless worlds, the presence of atmospheres on these planets has not been definitively ruled out \citep{Zieba_2023,Ih_2023,Lincowsky_2023,Turbet_2023,Ducrot_2023}. The first JWST transmission spectra of the TRAPPIST-1 planets, in general, show strong signs of stellar activity in the form of continuum variability, unocculted spots/faculae \citep{Lim_2023}, and flares \citep{Howard_2023}. Unocculted spots/faculae, responsible for the transit light source (TLS) effect \citep{Rackham_2018}, pose a significant challenge as they can mimic genuine atmospheric signals, greatly complicating the interpretation of transmission spectra. The TLS effect is a recurrent theme of numerous Cycle 1 and Cycle 2 transmission spectroscopy programs (e.g., \citealt{Lim_2023, Moran_2023, May_2023, Fournier-Tondreau_2024}); a fundamental problem that has yet to be solved.

LHS~1140 \citep{Dittmann_2017, Ment_2019, Lillo-Box2020, Cadieux_2024} is a keystone system for habitability studies. This M4.5 dwarf \citep{Dittmann_2017} hosts two small transiting planets: an outer planet LHS~1140\,b ($R_{\rm p} = 1.73$\,R$_{\oplus}$), a rare object in the middle of the radius valley \citep{Fulton_2017, Fulton_2018, Cloutier-Menou_2020} on a 24.7-day temperate orbit, and an inner planet LHS~1140\,c, a warm super-Earth ($R_{\rm p} = 1.27$\,R$_{\oplus}$) orbiting every 3.78\,days. The two planets receive 0.43 and 5.3 times the irradiation of Earth, respectively, with LHS~1140\,b comfortably situated within the Water Condensation Zone \citep{Turbet_2023}. \cite{Cadieux_2024} recently presented a joint study of almost all transit and radial velocity observations to date, obtaining a precision of 3\% on the masses and 2\% on the radii of LHS\,1140~b~and~c. Among the temperate rocky exoplanets, only those in the TRAPPIST-1 system are currently characterized to such precision. Their analysis of the composition of LHS~1140\,b shows that the planet could be enveloped in hydrogen ($\sim$0.1\% H/He by mass), akin to a smaller version of the temperate mini-Neptune K2-18\,b, which has a transmission spectrum characterized by prominent CH$_4$ and CO$_2$ absorption bands in the near-infrared \citep{Madhusudhan_2023}. Alternatively, LHS~1140\,b could be a water world with a surface layer of condensed water representing 9--19\% of the total mass. A 3D Global Climate Model (GCM) of the water world scenario predicts liquid water at the substellar point for a large range of atmospheric compositions \citep{Cadieux_2024}. The GCM also predicts a distinctive transmission spectrum featuring relatively small ($\sim$15\,ppm) CO$_2$ features at $\sim$2.8 and $\sim$4.3 $\mu$m.

Transmission spectroscopy with HST/WFC3 shows a tentative signal of H$_2$O/CH$_4$ near 1.4\,$\mu$m in a low mean molecular weight atmosphere \citep{Edwards_2021, Biagini_2024}, but a combination of unocculted spots and faculae offers an alternative explanation --- a hypothesis also supported by ground-based measurements \citep{Diamond-Lowe_2020}. These observations imply a large fractional coverage ($>$80\%) of stellar active regions which is somewhat surprising given that LHS~1140 is a relatively old ($>$5 Gyrs; \citealt{Cadieux_2024}), slowly rotating star ($P_{\rm rot} = 131 \pm 5$\,days; \citealt{Dittmann_2017}), with low levels of flaring activity \citep{Medina_2022}. Recent observations of two transits of LHS~1140\,b with JWST/NIRSpec (G235H/G395H, 1.7--5.2\,$\mu$m) ruled out an H$_2$-rich atmosphere and favored a high mean molecular weight atmosphere \citep{Damiano_2024}. Further TLS characterization with JWST is required, especially in the context of seeking the detection of small atmospheric signals that could be severely affected/biased by stellar activity.

This letter presents the results of a transmission spectroscopy program with JWST NIRISS/SOSS (0.6--2.8\,$\mu$m) aimed at both characterizing the stellar activity of LHS~1140 and providing a strong discriminating test of the mini-Neptune/water-world scenarios of \cite{Cadieux_2024}. We show that TLS is clearly detected in LHS~1140 and that a cloud-free mini-Neptune atmosphere for LHS~1140\,b is strongly excluded. 

The letter is structured as follows. The observations and data reduction steps are presented in Section~\ref{sec:observations} followed by the transit modeling in Section~\ref{sec:wlc} and spectral extraction analysis in Section~\ref{sec:spectrophotometric_fit}. The retrieved atmospheric properties of LHS~1140\,b and TLS constraints are presented and discussed in Section~\ref{sec:results} followed by a discussion of proposed follow-up observations in Section~\ref{sec:discussion}. We then conclude in Section~\ref{sec:conclusions}. 

\section{Observations} \label{sec:observations}

Two transits of LHS~1140\,b were obtained with JWST using the NIRISS instrument \citep{Doyon_2023} on UT2023-12-01 and UT2023-12-26 as part of program DD6543 (PI: Cadieux \& Doyon). The spectroscopic time series were carried using the SOSS mode \citep{Albert_2023}, offering a spectral resolution $R \approx 650$, with the SUBSTRIP256 subarray configuration to access the first two diffraction orders of NIRISS (0.6--2.8\,$\mu$m). Each visit comprised 949 integrations, covering 2.58\,hr before ingress, the 2.15-hour transit and 1.11\,hr after egress. The second visit fully captured a transit of LHS~1140\,c lasting 1.13\,hr and starting approximately 34\,min after first contact of LHS~1140\,b. Each integration was composed of $n_{\rm group}$ = 3 filling approximately 50\% of the full well to minimize charge migration and non-linearity effects \citep{Albert_2023}. We made use of two independent data reduction pipelines, namely \texttt{SOSSISSE} \citep{Lim_2023} and \texttt{supreme-SPOON} \citep{Feinstein_2023, Radica_2023, Radica_2024, Benneke_2024}, both yielding consistent transmission spectra with a mean (median) absolute deviation of 0.72$\sigma$ (0.60$\sigma$). More details of both pipelines and their comparison are provided in Appendix~\ref{appendix:pipeline_diff}. The \texttt{SOSSISSE} data products are hereafter used in the analysis. 

During the first visit, the star LHS~1140 was erroneously positioned outside the target acquisition field of view, causing a displacement of the orders on the detector with offsets $\Delta x = -157$\,px (dispersion) and $\Delta y = -12$\,px (cross-dispersion) relative to their nominal positions. This offset was calculated by comparing the pixel-to-wavelength calibrations derived from the two visits. A modified wavelength solution for the shifted spectral trace was determined through a cross-correlation between the stellar trace and a PHOENIX stellar atmosphere model \citep{Husser_2013}, following the procedure described by \cite{Feinstein_2023, Radica_2023}. The target acquisition failure is attributable to an incorrect epoch of proper motion entered during scheduling (LHS~1140 is a high proper motion target), which means this sequence was executed with blind guiding from the observatory. Fortunately, this data set can be fully recovered with a small wavelength shift in order 1 --- 0.70--2.67\,$\mu$m compared to 0.85--2.83\,$\mu$m during visit 2 --- and negligible change in order 2. 

The first visit is also affected by a background star\footnote{Gaia DR3 2371032989200665984, $G$\,=\,16.667, $d$\,=\,274.08\,pc, \citep{Gaia_2023}} that overlaps the spectral trace of LHS~1140 for wavelengths below 0.90\,$\mu$m in order 1, and between 0.60--0.75\,$\mu$m in order 2, contaminating the flux at the 1--2\% level. We estimate a spectral type M2.4$\pm$0.4V for the contaminant star using the $G-G_{\rm RP}$ calibration of \cite{Kiman_2019}. The $T_{\rm eff}$ estimate of 3516\,K from Gaia DR3 \citep{Gaia_2023} is also consistent with an M2V spectral type \citep{Pecaut_Mamajek_2013}. We constructed a model of the trace of the contaminant star from a M0V template (WASP-80; GTO 1201, PI: Lafrenière) observed with NIRISS. This template was aligned by maximizing the correlation coefficient between the derivatives in spatial direction $x$ of the template and observations in regions (pixels) where only the contaminant is seen. Once the optimal shift was determined, we computed the amplitude of the contamination pixel by pixel, and use the median of these amplitudes to scale the template to match the contaminant trace. We then subtracted this contaminant model from the images leaving flux residuals of the order of 0.1\% on the affected pixels of LHS~1140. Such residuals could decrease (dilute) the transit depth of LHS~1140\,b by 10\,ppm at most. This possible systematics is further mitigated in the next Sections by analysing both visits together, as visit 2 was not affected by a dispersed contaminant.

\section{White Light Curve Analysis} \label{sec:wlc}

One output of \texttt{SOSSISSE} is an amplitude time series $a(t)$ that corresponds to broadband flux measurements (details in Appendix~\ref{appendix:pipeline_diff}), hereafter referred to as the `white' light curve (WLC). The WLC is modeled with two components, one for transits and the other for treating systematic signals, both described below. The fit was performed using the \texttt{juliet} framework \citep{Espinoza_2019} that generates transit models with \texttt{batman} \citep{Kreidberg_2015} and implements Gaussian Process (GP) regression with \texttt{celerite} \citep{celerite_2017}. The posterior distribution is explored with nested sampling and 500 live points using the \texttt{dynesty} package \citep{Speagle_2020} available in \texttt{juliet}.

The transit component assumes circular orbits for both planets \citep{Gomes_2020}. The orbital parameters of planet $k$ ($k$: `b', `c') are the period $P_{k}$ fixed to the value from \cite{Cadieux_2024}, the time of inferior conjunction $t_{0,k}$, and the scaled semi-major axis $a_{k}/R_{\star}$ that we derive using a common stellar density $\rho_{\star}$. For $\rho_{\star}$, we use the same Gaussian prior as in \cite{Cadieux_2024} constructed from stellar mass and radius estimates. For all other parameters, we adopt wide uniform priors. The transits of LHS~1140\,b and c are parameterized with the planet-to-star radius ratio $R_{p, k}/R_{\star}$ and the impact parameter $b_{k}$. We fit for quadratic limb-darkening parameters $q_1$ and $q_2$ defined in \cite{Kipping_2013} that ensure physical solutions for values between 0 and 1. 

For the systematic component, we use GP regression to model two nuisance signals also reported in other NIRISS data sets (e.g., \citealt{Coulombe_2023, Radica_2024, Benneke_2024}): a beat pattern introduced by the observatory thermal control \citep{McElwain_2023} and correlated structures on the timescale of a few minutes suggestive of stellar granulation (e.g., \citealt{Kallinger_2014, Grunblatt_2017, Pereira_2019}). We use a combination of two simple harmonic oscillator (SHO) kernels in \texttt{celerite} to jointly model these systematic signals. We fix the period of the first SHO term to 204\,s, corresponding to the dominant periodicity in the out-of-transit fluxes based on its power (highest peak) in the generalized Lomb-Scargle periodogram \citep{Zechmeister_2009}. We fit for an amplitude $\sigma_{\rm beat}$ and a quality factor $Q_{\rm beat}$ for this oscillation. As in \citet{Radica_2024}, we use a critically damped SHO term ($Q$ fixed to $\sqrt{0.5}$) for the granulation-like signal and fit for an amplitude $\sigma_{\rm gra}$ and a timescale $\tau_{\rm gra}$ of this stochastic variation. An extra jitter term $\sigma_{\rm jitter}$ is added in quadrature to the diagonal of the covariance matrix to account for excess noise. We also fit for a baseline flux and a temporal slope for each visit. We measure slopes of $-0.26$ and $-0.36$ ppm per min during visits 1 and 2, respectively, a flux variation approximately 20--30 times weaker than observed during the transits of \hbox{TRAPPIST-1\,b} with NIRISS \citep{Lim_2023}. This is consistent with the fact that LHS~1140 is a slow-rotating star with a flux level remarkably stable over a timescale of a few hours, as also confirmed by TESS \citep{Ricker_2015}. 

The WLC of the two visits were jointly fit with a common $\rho_{\star}$, $q_1$, and $q_2$ to allow more precise constraints. The adopted priors and resulting posteriors (16$^{\rm th}$, 50$^{\rm th}$, and 84$^{\rm th}$ percentiles) of this fit are presented in Table~\ref{table:wlc_fit}. The best-fit transit and systematic components of the WLC are shown in the top panels of Figure~\ref{fig:spectral_fit}. Fitting visit 1 and visit 2 independently yielded consistent parameters within the 1$\sigma$ uncertainty, but we note a systematic increase of $\sim$40\,ppm in the transit depth of LHS~1140\,b during visit 2 due to the double transit. The joint fit alleviates the covariance between $R_{\rm p, b}$, $R_{\rm p, c}$, and $b_{\rm c}$ in visit 2 by using the information of a single transit of planet b from visit 1. This same argument is later used to extract combined transmission spectra of LHS~1140\,b and c by jointly fitting the two visits in each spectral channel.

\begin{deluxetable}{lcr}
\tablecaption{Transit and systematic parameters inferred from a joint fit of the white light curve of visits 1 and 2}
\tablehead{\colhead{Parameter} & \colhead{Prior} & \colhead{Posterior}}
\startdata
\multicolumn{3}{c}{\textit{Stellar parameters}} \\[0.1cm]
$\rho_{\star}$ (g\,$\cdot$\,cm$^{-3}$) & $\mathcal{N}\left(26.0, 2.6^2\right)$ & 26.7$^{+0.3}_{-0.6}$ \\
\multicolumn{3}{c}{\textit{Orbital parameters}} \\[0.1cm]
$P_{\rm b}$ (days) & (fixed)$^*$ & 24.73723 \\
$P_{\rm c}$ (days) & (fixed)$^*$ & 3.777940 \\
$t_{\rm 0,b}$ (\footnotesize{BJD\,-\,2460300}) & $\mathcal{U}\left(4.65, 4.75\right)$ & 4.69824(3) \\
$t_{\rm 0,c}$ (\footnotesize{BJD\,-\,2460300}) & $\mathcal{U}\left(4.65, 4.75\right)$ & 4.70775(4) \\
$a_{\rm b}/R_{\star}$ & (derived)$^{\dagger}$ & 95.2$^{+0.4}_{-0.7}$ \\
$a_{\rm c}/R_{\star}$ & (derived)$^{\dagger}$ & 27.1$^{+0.2}_{-0.3}$ \\
\multicolumn{3}{c}{\textit{Transit parameters}} \\[0.1cm]
$R_{\rm p,b}/R_{\star}$ (\%) & $\mathcal{U}\left(0, 50\right)$ & 7.439 $\pm$ 0.015 \\
$R_{\rm p,c}/R_{\star}$ (\%) & $\mathcal{U}\left(0, 50\right)$ & 5.312 $\pm$ 0.028 \\
$b_{\rm b}$ & $\mathcal{U}\left(0, 0.9\right)$ & 0.14$^{+0.05}_{-0.03}$ \\
$b_{\rm c}$ & $\mathcal{U}\left(0, 0.9\right)$ & 0.09$^{+0.06}_{-0.06}$ \\
$q_1$ & $\mathcal{U}\left(0, 1\right)$ & 0.16 $\pm$ 0.02 \\
$q_2$ & $\mathcal{U}\left(0, 1\right)$ & 0.22 $\pm$ 0.05 \\
\multicolumn{3}{c}{\textit{GP parameters}} \\[0.1cm]
$\sigma_{\rm beat}$ (ppm) & $\mathcal{LU}$(0.1, 1000) & 69 $\pm$ 6 \\
$Q_{\rm beat}$ & $\mathcal{LU}$(0.1, 1000) & 11.3$^{+3.4}_{-2.4}$ \\
$\sigma_{\rm gra}$ (ppm) & $\mathcal{LU}$(0.1, 1000) & 53 $\pm$ 5 \\
$\tau_{\rm gra}$ (min) & $\mathcal{LU}$(1, 100) & 5.2 $\pm$ 1.0 \\
$\sigma_{\rm jitter}$ (ppm) & $\mathcal{LU}$(0.1, 1000) & 95 $\pm$ 3 \\
\enddata
\tablecomments{$^*$Period from \cite{Cadieux_2024}. $^{\dagger}$Scaled semi-major axis ($a/R_{\star}$) derived from $\rho_{\star}$ and $P$}
\label{table:wlc_fit}
\end{deluxetable}

\section{Spectrophotometric Analysis} \label{sec:spectrophotometric_fit}

\begin{figure*}[t!]
\centering
\includegraphics[width=1\linewidth]{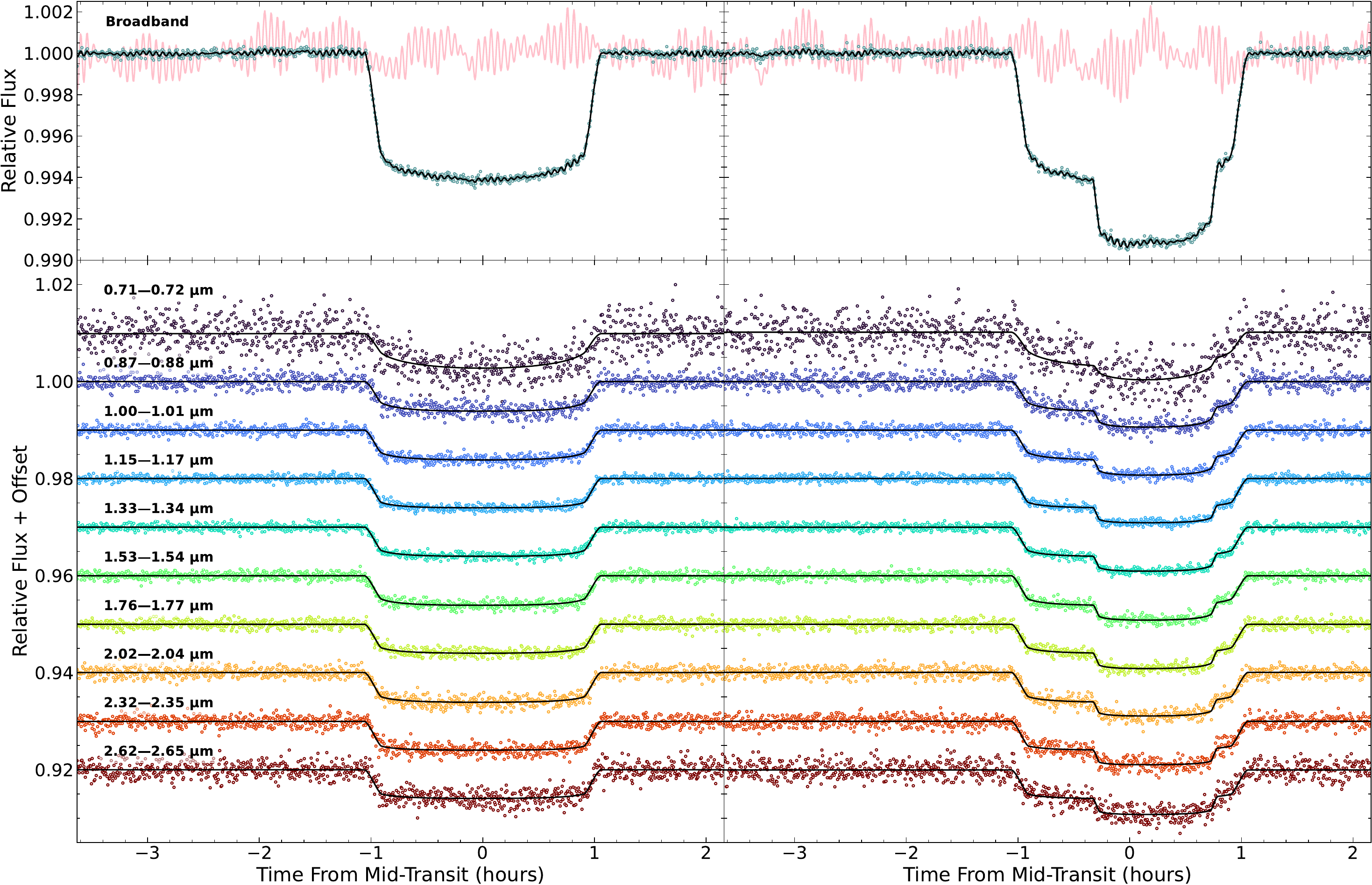}
  \caption{Spectrophotometric transit fits of LHS~1140\,b for visit 1 (UT2023-12-01, left) and visit 2 (UT2023-12-26, right) and LHS~1140\,c for visit 2 only. The broadband (`white') light curves are shown in top panels with the best-fit full model (transit + systematics) shown in black and systematics component only in pink (scaled 10$\times$ for clarity). For both visits, examples of 10 systematics-corrected spectroscopic bins are depicted with colored points in bottom panels.}
  \label{fig:spectral_fit}
\end{figure*}

We binned the extracted spectroscopic time series from \texttt{SOSSISSE} to $R \approx 100$ in both order 1 and order 2. We excluded wavelengths below (above) 0.86\,$\mu$m in spectral order 1 (2), as these wavelengths are already covered by the other order at a higher throughput. Because of the wavelength coverage differing between the two data sets (see Sect.~\ref{sec:observations}), this results in 145 spectrophotometric light curves for the first visit and 149 for the second visit. We also constructed a common wavelength grid at the native resolution corresponding to the average of the nearest pixels (in wavelength) between the two visits. This average grid deviates at most by 0.23\,nm from the wavelength calibrations from individual visits. We combined the spectral time series of visit 1 and visit 2 using the common wavelength grid, then binned to a $R \approx 100$ to end up with 142 joint spectrophotometric light curves. We then removed the systematic component determined from the WLC from all spectral light curves. At $R \approx 100$, the median photometric precision of the light curves is approximately 800\,ppm which prevents us from characterizing the beat pattern and granulation signals, both with amplitudes below 100\,ppm (Table~\ref{table:wlc_fit}).

To derive the transmission spectra of LHS~1140\,b and c, we fitted a transit model to each spectrophotometric light curve. Again, the fits were carried out with \texttt{juliet} using nested sampling and 500 live points for each spectral bin. We fixed $P_{k}$, $t_{0, k}$, $a_{k}/R_{\star}$, $b_{k}$ to the best-fit values from the WLC analysis (Table~\ref{table:wlc_fit}) and fit for $R_{\rm p,k}/R_{\star}$. The limb-darkening coefficients were calculated for all spectral channels with \texttt{ExoTic-LD} \citep{Grant_2022} using a custom PHOENIX model input ($T_{\rm eff}=3100$\,K, log~$g = 5$, [Fe/H] = 0.0) for LHS~1140. For the spectral fits, the corresponding $q_1$ and $q_2$ limb-darkening parameters are kept fixed. We allow for a baseline flux parameter and a $\sigma_{\rm jitter}$ for each visit and all spectroscopic bins. Examples of 10 spectroscopic channels jointly fitting the transits of LHS~1140\,b and c are shown in Figure~\ref{fig:spectral_fit}.

The final combined transit spectrum of LHS~1140\,b is presented in Figure~\ref{fig:transmission_spectrum} along with a binned version at $R\sim20$ to ease comparison with HST/WCF3 data. The spectrum is relatively flat above 1\,$\mu$m, but features a clear 200-ppm decrease towards shorter wavelengths. We show in the next section that this is a signature of the TLS effect from unocculted faculae. The HST/WFC3 data are fairly consistent with NIRISS albeit for one or two data points with larger transit depth blueward of 1.4\,$\mu$m which may have largely contributed to the tentative water/methane detection mentioned by \cite{Edwards_2021} and \cite{Biagini_2024}. The transmission spectra of LHS~1140\,b and c are available in Appendix~\ref{sec:appendix_spectra}.

\section{The Atmosphere of LHS~1140\,{\footnotesize{b}}} \label{sec:results}

We now use the combined transmission spectrum of LHS~1140\,b to jointly infer properties of the planetary atmosphere and stellar contamination from unocculted active regions resulting in stellar contamination (the transit light source, or `TLS' effect). Our analysis considers the evidence for stellar contamination (Sect.~\ref{sec:tls_only}), evidence for a planetary atmosphere through retrievals (Sect.~\ref{sec:retrieval}), and finally a comparison with GCM predictions of mini-Neptunes (Sect.~\ref{sec:forward_models}). We also provide a preliminary analysis of the transmission spectrum of LHS~1140\,c in Appendix~\ref{sec:appendix_lhs1140c} showing that its  spectrum is featureless and inconsistent with a cloud-free H$_2$-rich atmosphere.

\subsection{Transit Light Source Effect from LHS~1140} \label{sec:tls_only}

We first consider whether LHS~1140\,b's transmission spectrum can be explained only by unocculted stellar active regions. We modeled the TLS effect using the transmission spectra stellar contamination module of the \texttt{POSEIDON} retrieval package \citep{MacDonald_2017, MacDonald_2023}. We generate model spectra by multiplying a bare-rock transmission term, $(R_{\rm p}/R_{\star})^2$, by the wavelength-dependent stellar contamination factor from two distinct stellar heterogeneities (Equation 3, \citealt{Fournier-Tondreau_2024}). The active region spectra are calculated by interpolating PHOENIX stellar models using the \texttt{PyMSG} package \citep{Townsend_2023}. Our TLS model fits for the covering fractions and temperatures of the facula, spot, and quiet photosphere, yielding six free parameters: $R_{\rm p}$, $f_{\rm{fac}}$, $f_{\rm{spot}}$, $T_{\rm{fac}}$, $T_{\rm{spot}}$, and $T_{\rm{phot}}$. We fix the surface gravity of all active regions to $\log g = 5.04$ \citep{Cadieux_2024}. The priors are listed in Table~\ref{table:retrieval_results} and mainly follow those in \cite{Fournier-Tondreau_2024} for their two-heterogeneity model. Our \texttt{POSEIDON} model spectra are generated at $R = 20\,000$ from 0.60 to 2.9\,$\mu$m, convolved with a Gaussian kernel to the native resolution of NIRISS/SOSS ($R\approx650$), then multiplied by the instrument sensitivity function and binned down to the wavelength spacing of the data ($R \approx 100$). The parameter space is explored via the nested sampling package \texttt{PyMultiNest} \citep{Feroz_2008,Buchner_2014} with 2\,000 live points. The posterior distributions for this TLS-only model are shown in orange in the bottom row of Figure~\ref{fig:transmission_spectrum} (also in Appendix~\ref{sec:appendix_supplementary}), with the median and 1$\sigma$ confidence intervals listed in Table~\ref{table:retrieval_results}.

LHS~1140\,b's transmission spectrum shows definitive evidence of unocculted stellar faculae. We detect faculae at 5.8\,$\sigma$ confidence ($\Delta \ln \mathcal{Z} = 14.91$, where $\Delta \ln \mathcal{Z}$ is the increase in log Bayesian evidence for stellar contamination over a flat line). Our 6-parameter TLS model ($\chi^{2}_{\nu} = 1.73$, with 136 degrees of freedom), shown in Figure~\ref{fig:transmission_spectrum}, provides a significantly better fit compared to a flat line ($\chi^{2}_{\nu} = 1.96$, with 141 degrees of freedom). The reduced $\chi^2$ of 1.73 may be indicative of underestimated error bars, or perhaps imperfect TLS modeling due to inherent uncertainties in stellar atmosphere models \citep{Lim_2023, Jahandar2024}. We find unocculted faculae $72^{+84}_{-32}$\,K hotter than the photosphere covering $20^{+17}_{-12}$\% of the visible stellar disk. There is no evidence for spots. A wide range of covering fractions are consistent with the data due to the degeneracy between $f_{\rm fac}$ and $\Delta T = T_{\rm fac} - T_{\rm phot}$ (i.e., smaller/larger faculae need to be hotter/cooler to produce a similar spectral contamination). This level of stellar activity is consistent, within a factor of order unity, with the observed photometric peak-to-peak variability of $\sim$1\% of LHS~1140 \citep{Dittmann_2017} inferred from a 2-year monitoring campaign with MEarth \citep{Irwin_2009}.

We note that fitting a single stellar contamination model to LHS~1140b's combined spectrum amounts to assuming a similar surface distribution of unocculted heterogeneities during the times of the two visits separated by one orbital period of 24.7\,days, equivalent to $\sim$20\% of one stellar rotation. To validate this assumption, we also fit the same TLS-only model to the transmission spectrum of each individual visit, obtaining $f_{\rm fac} = 0.18^{+0.16}_{-0.11}$ and $T_{\rm fac} = 3151^{+77}_{-54}$\,K for visit 1 and $f_{\rm fac} = 0.06^{+0.16}_{-0.03}$ and $T_{\rm fac} = 3420^{+237}_{-227}$\,K for visit 2. We also observe consistent unocculted faculae properties from LHS~1140\,c's transmission spectrum with $f_{\rm fac} = 0.12^{+0.18}_{-0.07}$ and $T_{\rm fac} = 3293^{+205}_{-111}$\,K during visit 2 (Appendix~\ref{sec:appendix_lhs1140c}). This confirms that our two visits have consistent faculae properties, and hence it is valid to analyze the combined spectrum with a single TLS model. The consistent faculae parameters from our LHS~1140\,b and LHS~1140\,c analyses confirm that these unocculted faculae lie outside the stellar surface sampled by the two planetary transit chords. We additionally note that \cite{Damiano_2024} reported a spot crossing event during a different set of two transits of LHS~1140\,b observed with NIRSpec in July 2023. Their observations, conducted 124 days prior to our NIRISS data set (about one full rotation of LHS~1140), disfavor unocculted heterogeneities (their Table 4). However, the longer wavelengths covered by NIRSpec G235H and G395H are less sensitive to the TLS effect. Our analysis therefore critically underscore the need for NIRISS observations to characterize the nature and spectral impact of unocculted stellar active regions on transmission spectra.

\begin{figure*}[t!]
\centering
\includegraphics[width=\textwidth]{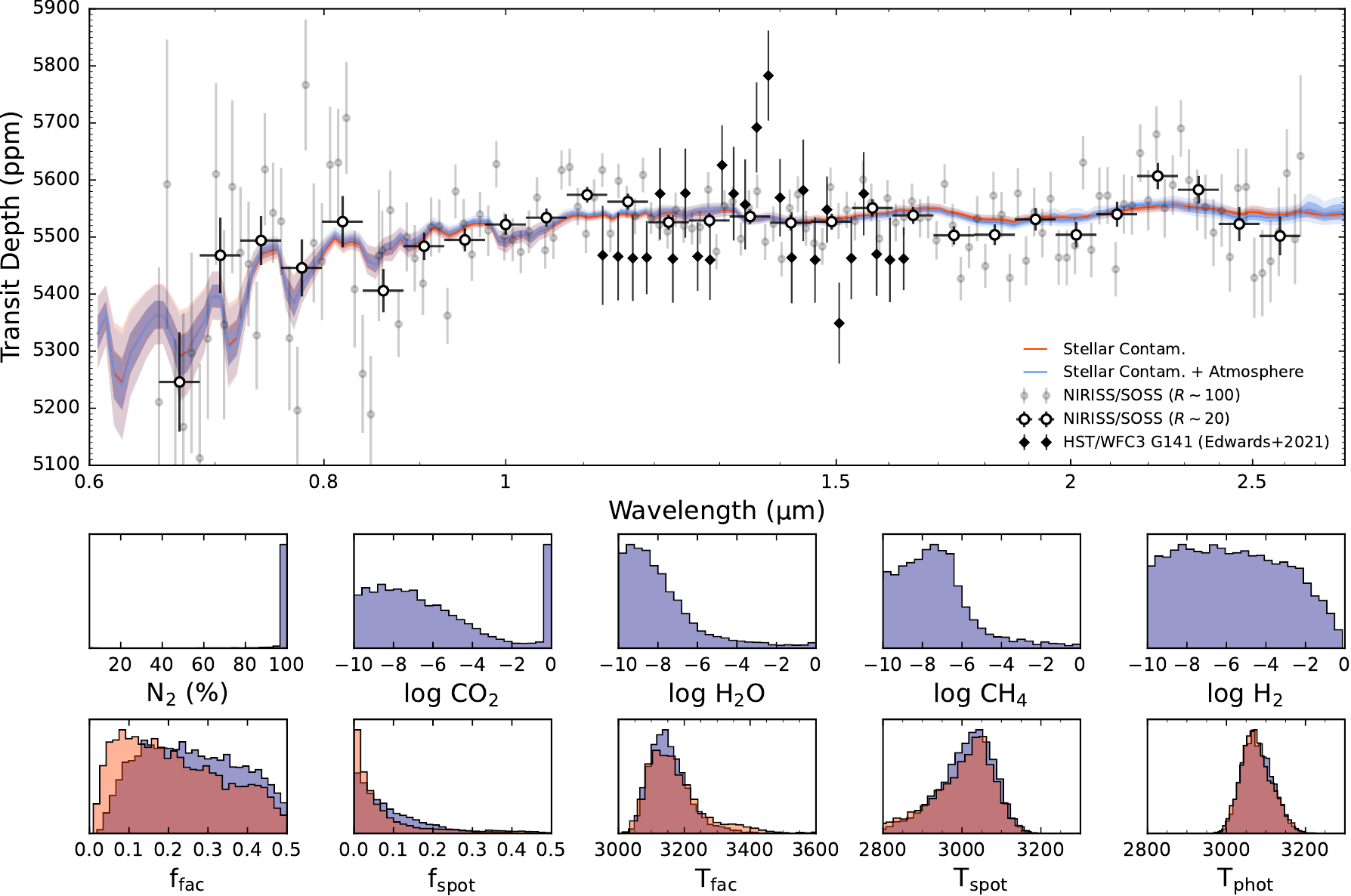}
  \caption{Combined transmission spectrum of LHS~1140\,b with NIRISS/SOSS from two transits and POSEIDON retrieval analysis results. The binned spectrum at $R\sim20$ (white points) and at a higher $R\sim100$ (gray points) significantly improve upon the HST/WFC3 data from \cite{Edwards_2021} (black diamonds). The median models with 1--2$\sigma$ confidence envelopes are shown for a stellar contamination model (orange) and a joint stellar contamination and planetary atmosphere model (blue), with their respective posterior distributions shown in the bottom panels (same colors). Confidence intervals for each parameter are given in Table~\ref{table:retrieval_results}. The transmission spectrum of LHS~1140\,b is mainly shaped by stellar contamination from unocculted faculae. A planetary atmosphere dominated by N$_2$ or CO$_2$ is compatible with the data, with the former pure composition being the maximum {\it a posteriori} model. A clear N$_2$-rich atmosphere combined with unocculted faculae is preferred at 2.3\,$\sigma$ over faculae alone (see Sect.~\ref{sec:retrieval} and Fig.~\ref{fig:TLS_Corrected_Transmission_Spectrum}).}
  \label{fig:transmission_spectrum}
\end{figure*}

\subsection{Atmospheric Retrieval} \label{sec:retrieval}

We next establish whether there is any evidence of an atmosphere on LHS~1140\,b. Given the presence of unocculted faculae established above, we performed a joint stellar contamination and planetary atmosphere retrieval analysis using \texttt{POSEIDON}. We consider a range of atmospheric retrieval models, including single-gas pure atmospheres, multiple gases, and atmospheres with or without aerosols. Our models consider the following gases (opacity sources in parentheses): CO$_2$ \citep{Tashkun2011}, H$_2$O \citep{Polyansky2018}, CH$_4$ \citep{Yurchenko2017}, H$_2$ (\citealt{Hohm1994} for Rayleigh scattering; \citealt{Karman2019} for collision-induced absorption), and N$_2$ (\citealt{Sneep2005} for Rayleigh scattering; \citealt{Karman2019} for collision-induced absorption). For multi-gas models, the volume mixing ratios follow a permutation-invariant centered-log-ratio (CLR) prior \citep{Benneke_2012}, which allows for any molecule to be equally likely \emph{a priori} to be the dominant gas. We assume isothermal atmospheres and fit for a reference radius, $R_{\rm p, ref}$, at a reference pressure of 1\,bar. We include a two-parameter power-law prescription for hazes \citep{MacDonald_2017} in all multi-gas atmospheric models. For the cloudy model, we assume an optically-thick gray opacity/surface with all layers deeper than $P_{\rm surf}$ set to infinite opacity. We allow $M_{\rm p}$ to vary as a free parameter following a Gaussian prior ($5.60 \pm 0.19 M_{\oplus}$; \citealt{Cadieux_2024}). For the stellar contamination, we use the same TLS parameterization as outlined above (i.e., 5 parameters to fit for both faculae and spots). We used 2\,000 \texttt{PyMultiNest} live points for all retrievals. A summary of the free parameters and their priors is provided in Table~\ref{table:retrieval_results}.

Our retrieval analysis identifies tentative evidence for a cloud-free N$_2$-dominated atmosphere on LHS~1140\,b (2.3$\sigma$). We first considered models with a single-gas atmosphere with negligible cloud opacity in the upper atmosphere (all including stellar contamination), finding that the fit did not improve for CO$_2$, H$_2$O, CH$_4$, or H$_2$ pure atmospheres. For example, our 100\% CO$_2$ model has a lower Bayesian evidence compared to stellar contamination only ($\ln \mathcal{Z} = 1141.02$ vs.\ 1142.15) --- reflecting the `Occam penalty' from adding one or more parameters that do not improve the fit \citep[see e.g.,][]{Trotta2008} --- since there are no apparent CO$_2$ features in our NIRISS spectrum. Pure CO$_2$ atmospheres must have $T < 233$\,K (2$\sigma$ upper limit) to be compatible with our non-detection of CO$_2$ absorption (i.e., high-temperature pure CO$_2$ atmospheres are ruled out). However, our 100\% N$_2$ model has the highest Bayesian evidence of all our models ($\ln \mathcal{Z} = 1144.65$), with a $\Delta \ln \mathcal{Z} = 2.50$ (2.8$\sigma$) for N$_2$ + stellar contamination over stellar contamination alone. The evidence for N$_2$ arises from a combination of a residual short-wavelength slope compatible with N$_2$ Rayleigh scattering and a weak feature near 2.2\,$\mu$m attributable to N$_2$--N$_2$ collision-induced absorption (CIA) --- see Appendix~\ref{sec:appendix_supplementary} and Figure~\ref{fig:TLS_Corrected_Transmission_Spectrum}. Our multiple gas retrieval reaffirms that N$_2$ is the favored background gas, even when hazes are included in the model, with a retrieved abundance of $100^{+0}_{-2}$\% (see Fig.~\ref{fig:transmission_spectrum}, blue posteriors), indicating that our data rule out the tail to lower N$_2$ abundances present in the CLR prior. We rule out cloud-free H$_2$O-rich, CH$_4$-rich, and H$_2$-rich atmospheres, with 2$\sigma$ upper limits of $\log \mathrm{H_2 O} < -2.94$, $\log \mathrm{CH_4} < -2.78$, and $\log \mathrm{H_2} < -0.95$ ($<$ 11\%), respectively. While our multiple gas retrieval does allow for cold ($T \approx 100$\,K) CO$_2$-rich atmospheres (corresponding to a relatively featureless atmospheric contribution to the spectrum), this solution is not statistically favored, as discussed above. To provide a more conservative estimate for the N$_2$ detection significance, we calculated the Bayesian evidence for a multiple gas model including all the other gases and hazes but without N$_2$ (ensuring the possibility of both CO$_2$-dominated atmospheres and hazes are propagated into the Bayesian evidence), finding a conservative $\Delta \ln \mathcal{Z} = 1.50$ (2.3$\sigma$) in favor of an N$_2$-dominated atmosphere. Our tantalizing evidence for a N$_2$-dominated atmosphere on a habitable zone super-Earth raises a resounding call for additional observations of LHS~1140\,b to confirm this result.

We additionally considered the potential for aerosols in LHS~1140\,b's atmosphere. First, we reiterate that our multi-gas retrieval discussed above included a parametric prescription for atmospheric hazes in the form of a power law \citep{MacDonald_2017}. The haze parameters are unconstrained in all our retrievals (see posteriors in Appendix~\ref{sec:appendix_supplementary}). The reason
for this is that hazes still exist within an atmosphere, providing an additional opacity enhancing the dominant Rayleigh slope,
but the retrieval must first ‘select’ a background gas before the impact of a haze can be established. If N$_2$ is the background gas, N$_2$ Rayleigh scattering alone provides a good fit, and additional haze opacity is not needed (hazes are redundant parameters). If CO$_2$ is the dominant gas, then hazes have no spectral effect due to the high molecular weight and low temperature shrinking the scale height and hence flattening the spectrum. Therefore, allowing for hazes does not affect the inference of a clear N$_2$-dominated atmosphere as the favored model. When we include a cloud/surface in our multi-gas retrieval we obtain a bimodal solution: (i) a clear $\sim$100\% N$_2$ atmosphere, as above, or (ii) an optically-thick cloud deck at low $P_{\rm surf} \sim 10^{-5}$\,bar with no constraints on the atmospheric composition (i.e., such high-altitude clouds would render any transmission spectrum featureless, regardless of the atmospheric composition). The Bayesian evidence decreases when adding a cloud deck (see Table~\ref{table:retrieval_results}), indicating that this is a redundant parameter not required to explain LHS~1140\,b's transmission spectrum. We also stress that such exceedingly high-altitude clouds are probably unrealistic. According to 3D GCMs of LHS~1140\,b for both the water world \citep{Cadieux_2024} and hydrogen-rich scenarios (see below), a cloud deck (e.g., H$_2$O, CO$_2$ clouds) should form much deeper in the atmosphere ($P \approx1$\,bar). Therefore, while our present observations do not strongly favor a clear N$_2$-dominated atmosphere over a high-altitude cloud deck, the latter scenario is less plausible from a GCM standpoint.  

We examine the evidence for N$_2$ in the atmosphere of LHS~1140\,b and present our full atmospheric retrieval results and posterior distributions in Appendix \ref{sec:appendix_supplementary}.

\subsection{3D GCM Forward Modeling} \label{sec:forward_models}

\begin{figure*}[t!]
\centering
\includegraphics[width=1\linewidth]{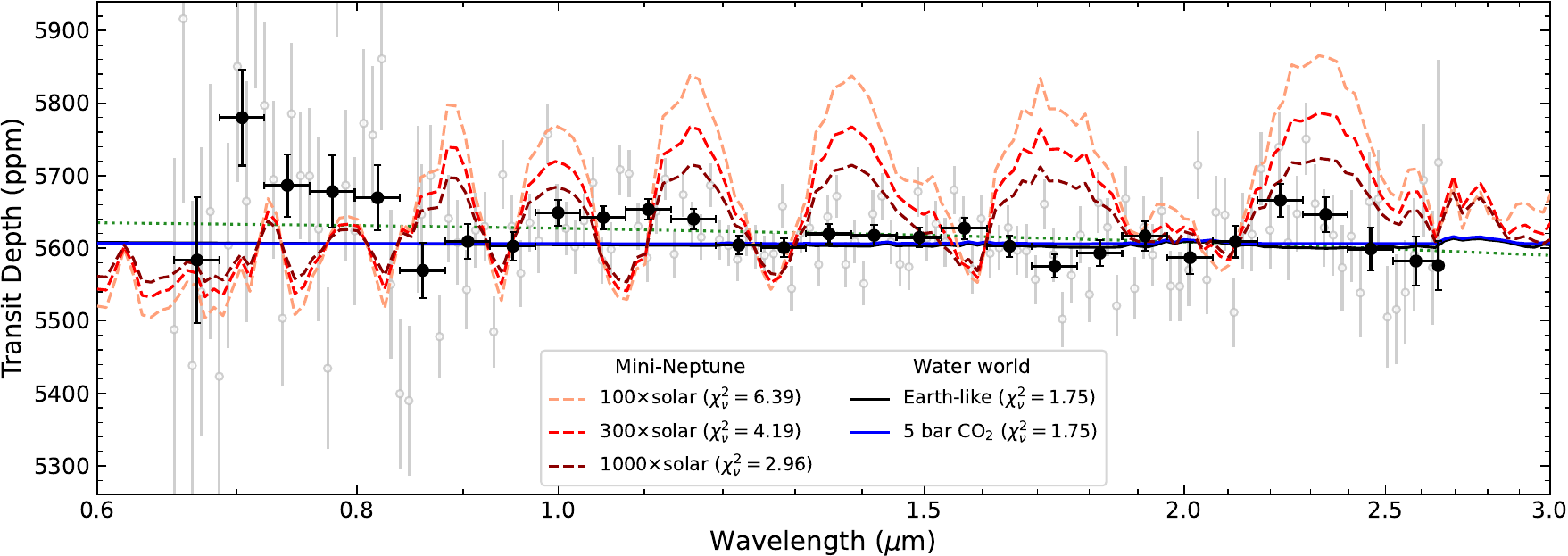}
  \caption{Stellar contamination-corrected combined spectrum of LHS~1140\,b using the best-fit solution of Figure~\ref{fig:transmission_spectrum}. The black points are binned to $R\sim20$ while the gray points are at a higher $R\sim100$. Dashed lines depict GCM-based forward models for mini-Neptune atmosphere (100$\times$, 300$\times$, 1000$\times$solar metallicity) and solid lines the water world GCMs of \cite{Cadieux_2024} for an Earth-like atmosphere (1\,bar N$_2$, 400\,ppm CO$_2$) and a pure CO$_2$ atmosphere at 5\,bar. An H$_2$-rich atmosphere for LHS~1140\,b is formally rejected ($>$10$\sigma$) by the NIRISS/SOSS data. A flat solution (e.g., water world or airless planet) provides a significantly better fit but the reduced $\chi^2$ of 1.75 may indicate an underestimation of error bars, an imperfect TLS correction, or the presence of residual spectral features (e.g., slope). An apparent linear trend with a slope of approximately 40\,ppm across the NIRISS domain (dotted green line) improve the fit ($\chi^2_{\nu} = 1.72$), but is not predicted by our GCMs.}
  \label{fig:forward_models}
\end{figure*}

Here we compare the observed spectrum with realistic predictions from 3-D GCMs of a mini-Neptune with a 80\,bar H$_2$-rich atmosphere of various compositions: 100$\times$, 300$\times$, and 1000$\times$solar metallicity. For this, we used the Generic PCM (Planetary Climate Model), historically known as the Generic LMD GCM, as in \cite{Cadieux_2024}. The model is capable of simulating various types of exoplanets, including terrestrial planets in the TRAPPIST-1 system (\citealt{Turbet_2018}; \citealt{Fauchez_2019}) and sub-Neptunes such as GJ\,1214\,b \citep{Charnay_2015} or K2-18\,b \citep{Charnay_2021}. The GCM simulations performed here closely follow the methodology described in \citet{Charnay_2021}, except that star and planet properties have been adapted to LHS~1140\,b. The GCM experiments are described in more details in Appendix~\ref{sec:appendix_gcm}.

Following the methodology of \cite{Fauchez_2019}, the outputs of the GCMs are used as inputs to the Planetary Spectrum Generator (\citealt{Villanueva_2018}) to generate synthetic transmission spectra using the GlobES module \citep{Villanueva_2022} that consider the 3D nature of the atmosphere and cloud in a self-consistent manner. The simulations include atmospheric refraction, collision-induced absorption (CIA) and multiple scattering. Irrespective of atmospheric composition, all forward models have a cold trap between $\sim$0.1 and 1\,bar (see Fig.~\ref{fig:PTprofiles_GCM}), favoring condensation of H$_2$O into clouds. For these realistic cloudy mini-Neptune models, CH$_4$ and CO$_2$ (to a lesser extent) are the only observable transmission features in the near-infrared. Figure~\ref{fig:forward_models} compares the TLS-corrected transmission spectrum of LHS~1140\,b with these updated mini-Neptune models along with water world GCMs taken from \cite{Cadieux_2024} for an Earth-like atmosphere (1\,bar N$_2$, 400\,ppm CO$_2$) and a pure CO$_2$ atmosphere at 5\,bar. All hydrogen-rich GCMs are formally rejected by the data (fitting for an offset) with a confidence level exceeding 10\,$\sigma$, with a $\Delta \ln \mathcal{Z} = -326.90$, $-172.22 $, $-84.98$ for the 100$\times$, 300$\times$, and 1000$\times$solar metallicity models, respectively, compared to a flat spectrum. The water world GCMs or a flat line (airless planet) provide a much better fit to the data. However, we note an apparent slope of $\sim$40\,ppm in the spectrum that is not predicted by the GCMs explored in this study. The possible origin of this residual slope is discussed in Appendix~\ref{sec:appendix_supplementary}.

\section{Discussion} \label{sec:discussion}

\subsection{A Secondary Atmosphere on LHS~1140\,b?} \label{sec:secondary_atm}

While an exceptionally cloudy mini-Neptune is compatible with our NIRISS spectrum, such a scenario would be in stark contrast with the relatively clear atmospheres predicted by self-consistent GCMs of LHS~1140\,b. Recent observations of the temperate mini-Neptunes K2-18\,b \citep{Madhusudhan_2023} and TOI-270\,d \citep{Benneke_2024} have detected strong spectral features associated with H$_2$-rich atmospheres, with relatively little cloud opacity. Instead, our tentative inference of an N$_2$-dominated atmosphere suggests LHS~1140\,b is a water world with a secondary atmosphere \citep{Forget_2014,Kite_2018,Kite_2020,Marounina_2022}.

One line of evidence supporting atmospheric retention on LHS 1140\,b is that the planet is likely resilient to atmospheric mass loss of light elements (H$_2$, He) from processes such as XUV-driven photoevaporation \citep{Owen_2017} and core-powered mass loss \citep{Ginzburg_2018}. According to \citealt{Cadieux_2024} (Figure 4, therein), an initial envelope mass fraction as low as $\sim$0.1\% H/He could mostly persist after 10\,Gyr thanks to the relatively low instellation of LHS~1140\,b combined with its relatively high surface gravity. Based on these results, it is probable that an atmosphere composed of any elements heavier than hydrogen or helium would also be retained by the planet over similar timescales.

\begin{figure}[h!]
\centering
\includegraphics[width=1\linewidth]{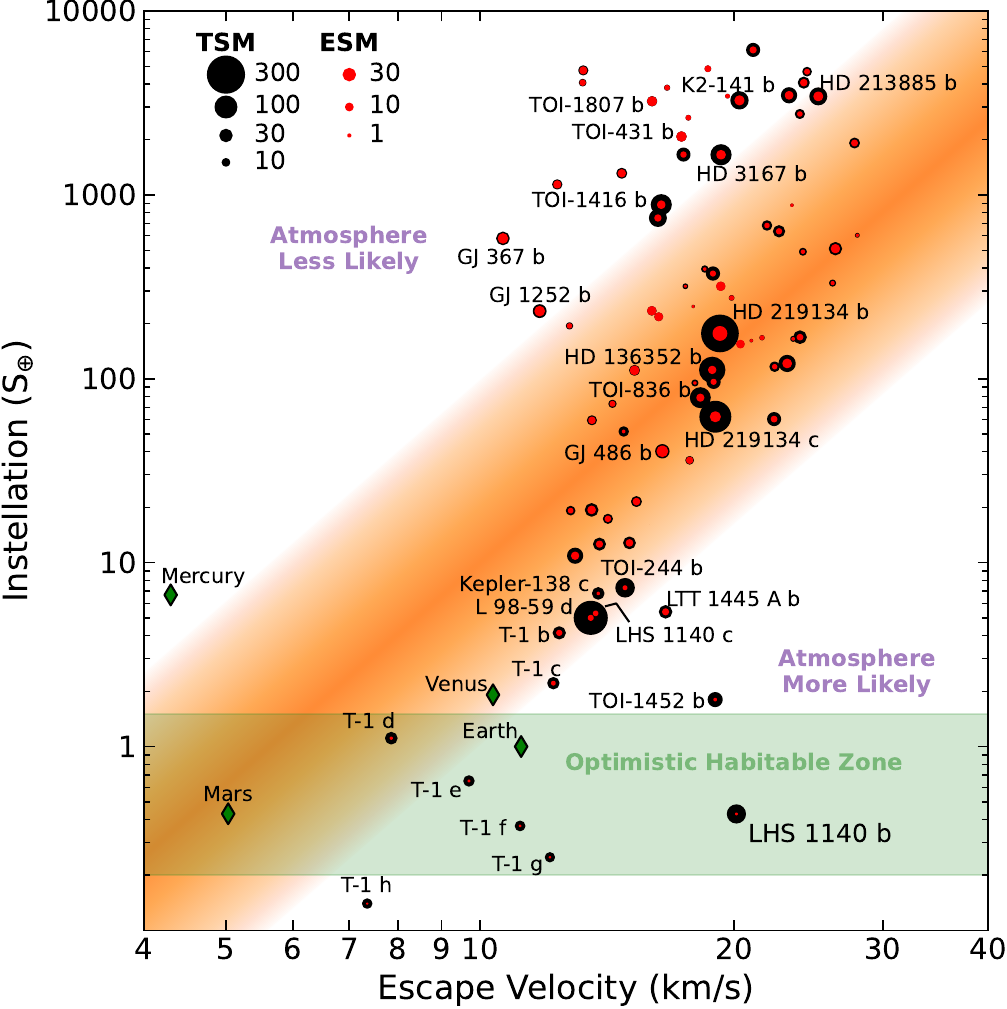}
  \caption{Escape velocity--instellation diagram of small exoplanets ($R_{\rm p} < 1.8$\,R$_{\oplus}$). The empirical ``cosmic shoreline'' ($S \propto v_{\rm esc}^4$) of \cite{Zahnle_2017} based on Solar System bodies is depicted in orange with an arbitrary envelope to underscore uncertainty with spectral types. The size of data points scale with their Transmission and Emission Spectroscopy Metrics \citep{Kempton_2018} to highlight favorable targets for JWST atmospheric characterization. Targets with $S < 10$\,S$_{\oplus}$ or with TSM/ESM above 90$^{\rm th}$ percentile are annotated. The TRAPPIST-1 system is abbreviated as ``T-1''. The optimistic habitable zone (Early Mars/Recent Venus) defined by \cite{Kopparapu_2013} is shown in green (approximated for a range of spectral types). With a relatively high escape velocity and low instellation, LHS~1140\,b likely retained an atmosphere.}
  \label{fig:shoreline}
\end{figure}

Another way to present this argument is illustrated in Figure~\ref{fig:shoreline}, showing the instellation vs.\ the escape velocity of LHS~1140\,b ($S = 0.43 \pm 0.03$\,S$_{\oplus}$, $v_{\rm esc} = 20.13\pm0.37$\,km/s; \citealt{Cadieux_2024}) along with other exoplanets smaller than 1.8\,R$_{\oplus}$ using the NASA Exoplanet Archive \citep{Akeson_2013}. This radius cut is chosen to focus on terrestrials, super-Earths, and potentially water worlds, excluding exoplanets predominantly composed of gas. The empirical ``cosmic shoreline'' ($S \propto v_{\rm esc}^4$) of \cite{Zahnle_2017} based on Solar System objects appears as an orange band in the figure, a region that separates planets with an atmosphere from those without. Reconnaissance atmospheric characterization in transmission and/or in emission with JWST of a few key targets will be required to establish this shoreline across spectral types. As shown in this diagram, LHS~1140\,b, TOI-1452\,b \citep{Cadieux_2022}, TRAPPIST-1\,f and TRAPPIST-1\,g \citep{Agol_2021} appear as the most favorable small exoplanets to host an (secondary) atmosphere.

\subsection{Future observations} \label{sec:future}

Considering the significant implications associated with the potential detection of a secondary atmosphere on LHS~1140\,b, seeking unambiguous evidence warrants a combined approach of eclipse and transmission spectroscopy. Here, we discuss the required observations to ascertain with high certainty the presence of a secondary atmosphere on LHS~1140\,b (or lack thereof). It should be noted that due to the system's limited visibility with JWST, only 4 or 5 transits/eclipses of LHS~1140\,b are observable in a given year.

\subsubsection{Eclipse Photometry}

Eclipse photometry centered on the 15-$\mu$m CO$_2$ absorption feature is a powerful tool for constraining the presence of an atmosphere on small rocky planets, e.g., TRAPPIST-1\,b \citep{Greene_2023} and TRAPPIST-1\,c \citep{Zieba_2023}.  The eclipse depth is given by $F_{\rm p}/F_\star$ where $F_{\rm p}$ and $F_\star$ are respectively the flux from the dayside of the planet and the star during the eclipse. Assuming a BT-COND stellar atmosphere model \citep{Allard_2012} with an effective temperature of 3100 K anchored to the 2MASS $J$ magnitude of LHS~1140, we find $F_\star=10.1$\,mJy for the F1500W MIRI filter. The observations of TRAPPIST-1\,b at 15\,$\mu$m suggest that the stellar model prediction for $F_{\star}$ should be accurate to $\sim$10\% \citep{Greene_2023}. A blackbody is assumed for $F_{\rm p}$ with the equilibrium temperature $T_{\rm eq} = T_{\rm eff} \ \sqrt{R_\star/a} \ \left[f(1-A_{\rm B})\right]^{1/4}$, where $T_{\rm eff}$ is the effective temperature of the host star, $R_\star$ its radius, $a$ the planet semi-major axis, $A_{\rm B}$ the planet Bond albedo and $f$ the re-radiation factor that can take two extreme values: $f=1/4$ corresponding to a homogeneous distribution of the energy across the planet and $f=2/3$ corresponding to no heat redistribution to the night side \citep{Esteves2013}. 

\begin{figure}[ht!]
\centering
\includegraphics[width=1\linewidth]{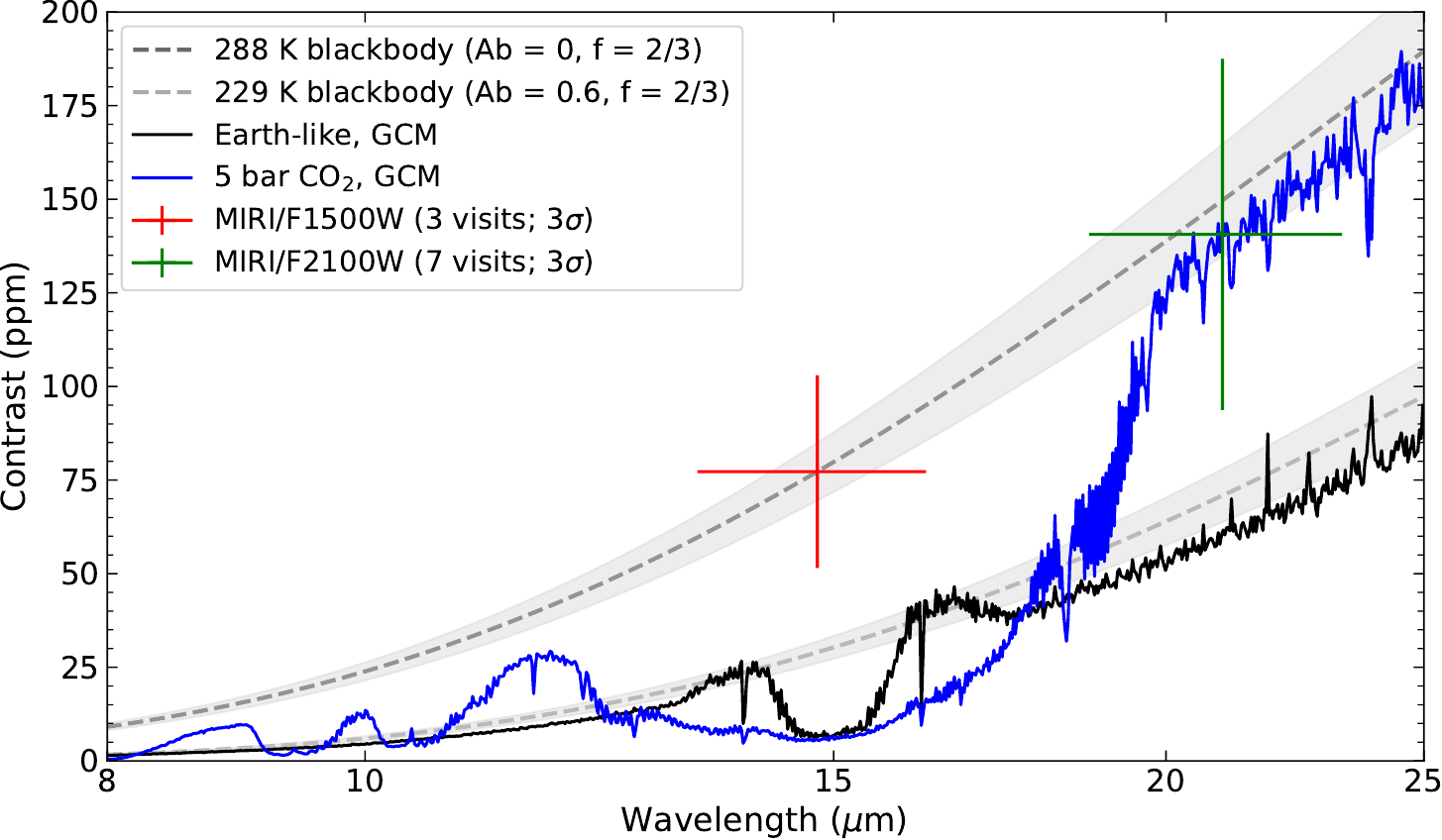}
  \caption{Predicted emission spectrum of LHS~1140\,b for various scenarios including the Earth-like (1\,bar N$_2$, 400 ppm CO$_2$) and pure CO$_2$ (5\,bar) atmosphere cases from the GCMs of \cite{Cadieux_2024}. Extreme Bond albedos cases (0 or Europa-like) with no atmospheric heat redistribution ($f$\,=\,2/3) are shown with dashed lines. The grey enveloppe represents the 10\% uncertainty of LHS~1140's stellar flux inferred from a BT-COND model. The red cross, centered on the deepest airless case ($A_{\rm B}$\,=\,0, $f$\,=\,2/3), gives the predicted uncertainty at 15\,$\mu$m for three visits needed to reach a 3-$\sigma$ detection assuming photon-limited performance. Similarly, the green cross gives the number of visits needed (seven) for detecting the deepest eclipse at 21\,$\mu$m. Each visit is assumed to last $\sim$7.5 hours (1\,hr + 3 times the transit duration), a similar experimental design adopted by \cite{Greene_2023} and \cite{Zieba_2023} for the MIRI observations of TRAPPIST-1\,b and c, respectively.} \label{fig:eclipse_depth}
\end{figure}

Figure~\ref{fig:eclipse_depth} shows the predicted emission spectrum of LHS~1140\,b for two airless models with different albedos and the two water world GCM cases of \cite{Cadieux_2024}. Three visits with MIRI would be sufficient to detect/exclude at the 3\,$\sigma$ level the most optimistic scenario ($A_{\rm B}=0, f=2/3$) corresponding to a dark, airless planet which is the coreless case inferred from the internal structure model of \cite{Cadieux_2024}. An airless Europa-like surface ($A_{\rm B}=0.6$, \citealt{Buratti1983}) would require $\sim$19 visits to detect the secondary eclipse, i.e., observing every eclipse for nearly 5 years. Detecting and discriminating the water world cases (an Earth-like or CO$_2$-rich atmosphere) is out of reach through eclipse photometry at 15\,$\mu$m only, and would probably require the combination of many filters (e.g., 15 + 21\,$\mu$m) over many eclipses.

\subsubsection{Transmission spectroscopy}

\citet{Cadieux_2024} estimated that 12 transits (six times each of NIRSpec G235 and G395) would be required to detect the CO$_2$ features longward of 2.8 $\mu$m, an exposure time in line with the nine additional G395 visits proposed by \citealt{Damiano_2024} to detect the CO$_2$ feature at 4.3 $\mu$m. While G395 may be the most attractive mode for maximizing the signal-to-noise of the 4.3 $\mu$m feature, it comes with the significant scientific compromise that the atmospheric water abundance in the presence of TLS, as observed in LHS~1140\,b, is likely to be poorly constrained. The same ambiguity issue between TLS and water detection associated with the G395 mode alone has been noted by other NIRSpec programs on warm super-Earths (e.g., \citealt{Moran_2023}; \citealt{May_2023}). Moreover, the G395 mode is insensitive to short wavelength spectral slopes associated with hazes and Rayleigh scattering. Our observations provide tantalizing evidence (2.3\,$\sigma$) for a Rayleigh scattering slope and a CIA feature around 2.2 $\mu$m that could be explained by a N$_2$-rich atmosphere. Such a signal could be confirmed with approximately 4 additional visits that should also include NIRSpec observations since the N$_2$-N$_2$ CIA shows stronger opacity at 4.3\,$\mu$m which is very close to the CO$_2$ feature.  Clearly, the optimal experimental design for exploring the full diversity of secondary atmospheres (N$_2$- or CO$_2$-rich) along with proper TLS characterization calls for the widest possible wavelength coverage as offered by the combination of NIRISS/SOSS and NIRSpec/G395.  Observing within the same season (July or December) through two consecutive visits between NIRISS and NIRSpec minimizes the change in stellar rotation phase which is only $\sim$20\% given the relative orbital period of LHS~1140\,b (24.7\,days) and the stellar rotation period of $\sim$131\,days. The two consecutive transits of LHS~11140\,b observed with NIRISS that we presented in this letter is the proof of concept of this experimental design, showing that the TLS can vary weakly within one observing season, implying that the G395 and NIRISS data acquired within one season could be analyzed jointly under the reasonable assumption that both visits share a similar stellar heterogeneity configuration. 

\section{Summary \& Conclusion} \label{sec:conclusions}

We have presented the 0.65--2.7\,$\mu$m transmission spectrum of the temperate planet LHS~1140\,b obtained from two visits with JWST/NIRISS, the second of which coincidentally involved a transit of LHS~1140\,c. All spectra exhibit a low level of stellar contamination caused by unocculted faculae with covering fractions at the level of $\sim$20\%. GCMs of mini-Neptunes with various atmospheric compositions (100$\times$, 300$\times$, 1000$\times$solar metallicity) are all excluded with a significance greater than 10\,$\sigma$. A spectral retrieval analysis also excludes a clear H$_2$-rich atmosphere, with the most likely atmospheric scenario being that of an N$_2$- or CO$_2$- dominated envelope. The former N$_2$-rich atmospheric composition is favored by the data at 2.3\,$\sigma$ from a tentative detection of N$_2$ Rayleigh scattering combined with weak N$_2$-N$_2$ CIA absorption near 2.2\,$\mu$m. While an H$_2$-rich atmosphere with a relatively high cloud deck is consistent with the NIRISS observations, such a muted spectrum is in contradiction with cloud formation simulated in GCMs and the relatively clear atmospheres detected on the temperate mini-Neptunes K2-18\,b \citep{Madhusudhan_2023} and TOI-270\,d \citep{Benneke_2024}. 

The NIRISS observations are more likely consistent with the water world scenarios presented by \cite{Cadieux_2024}, with or without a secondary atmosphere, a conclusion also supported by recent transmission spectroscopy data obtained with JWST/NIRSpec \citep{Damiano_2024}. Of all nearby small temperate planets, LHS~1140\,b is the most likely to have retained a secondary atmosphere based on its low instellation and high surface gravity. The next obvious step to better constrain LHS~1140\,b's atmospheric composition is to perform a joint analysis of both NIRISS and NIRSpec datasets, ideally with a common data reduction methodology. 

Only three visits of 15\,$\mu$m eclipse photometry with MIRI is required to yield a 3-$\sigma$ detection of the secondary eclipse associated with a dark, airless planet, but nearly five years (4--5 eclipses per year) is required if its albedo is closer to an icy surface ($A_{\rm B}\sim0.5$) as suggested by the relatively high water mass fraction ($\sim$15\%) of LHS~1140\,b. Transmission spectroscopy is the most efficient method to detect the potential secondary atmosphere of LHS~1140\,b through consecutive observations between NIRISS/SOSS and NIRSpec/G395 during the same season to allow for a proper characterization of stellar contamination at all epochs and explore N$_2$- and CO$_2$-dominated atmospheres. Given the limited visibility of LHS~1140\,b, several years worth of observations may be required to detect its potential secondary atmosphere, a program that should be initiated as soon as possible given the limited lifetime of JWST. LHS~1140\,b is arguably the best temperate transiting planet for which liquid surface water may be indirectly inferred through the detection of sufficient level of atmospheric CO$_2$.

\fakesection{Acknowledgments}
\vspace{1cm}
We thank the anonymous referee for a very thoughtful report whose constructive comments and suggestions improved the clarity and quality of the paper. This work is based on observations made with the NASA/ESA/CSA JWST. The data were obtained from the Mikulski Archive for Space Telescopes at the Space Telescope Science Institute, which is operated by the Association of Universities for Research in Astronomy, Inc., under NASA contract NAS 5-03127 for JWST. These observations are associated with program \#6543 (\href{doi.org/10.17909/yf93-zf91}{10.17909/yf93-zf91}). This work is partly supported by the Natural Science and Engineering Research Council of Canada, the Canadian Space Agency and the Trottier Institute for Research on Exoplanets (iREx) through the Trottier Family Foundation. This work benefited from support of the Fonds de recherche du Québec – Nature et technologies (FRQNT), through the Center for Research in Astrophysics of Quebec. R.J.M. is supported by NASA through the NASA Hubble Fellowship grant HST-HF2-51513.001, awarded by the Space Telescope Science Institute, which is operated by the Association of Universities for Research in Astronomy, Inc., for NASA, under contract NAS 5-26555. C.P.-G. acknowledges support from the NSERC Vanier scholarship, and iREx. M.T. thanks the Gruber Foundation for its generous support to this research, support from the Tremplin 2022 program of the Faculty of Science and Engineering of Sorbonne University, and the Generic PCM team for the teamwork development and improvement of the model. This work was performed using the High-Performance Computing (HPC) resources of Centre Informatique National de l'Enseignement Supérieur (CINES) under the allocations No. A0120110391 and A0140110391 made by Grand Équipement National de Calcul Intensif (GENCI). M.F.T. acknowledges financial support from the Clarendon Fund and the FRQNT.

\vspace{5mm}
\facilities{JWST/NIRISS}

\software{\texttt{Astropy} \citep{Astropy_2018}; \texttt{matplotlib} \citep{Hunter_2007}; \texttt{juliet} \citep{Espinoza_2019}; \texttt{batman} \citep{Kreidberg_2015}; \texttt{SciPy} \citep{Virtanen_2020}; \texttt{NumPy} \citep{Harris_2020}; \texttt{PSG} \citep{Villanueva_2022}; \texttt{SOSSISSE} (\href{https://github.com/njcuk9999/sossisse}{\footnotesize \texttt{github.com/njcuk9999/sossisse}}); \texttt{supreme-SPOON} (\href{https://github.com/radicamc/supreme-spoon}{\footnotesize \texttt{github.com/radicamc/supreme-spoon}}); \texttt{exofile} (\href{https://github.com/AntoineDarveau/exofile}{\footnotesize \texttt{github.com/AntoineDarveau/exofile}})}

\bibliography{LHS1140b}{}
\bibliographystyle{aasjournal}

\appendix
\counterwithin{table}{section}
\counterwithin{figure}{section}

\section{Data Reduction Pipelines} \label{appendix:pipeline_diff}
\setcounter{figure}{0}
\renewcommand{\thefigure}{A\arabic{figure}}
\setcounter{table}{0}
\renewcommand{\thetable}{A\arabic{table}}

The NIRISS data were reduced with two independent pipelines (\texttt{SOSSISSE} and \texttt{supreme-SPOON}) briefly described here.

The \texttt{SOSSISSE} package starts with the \textit{rateints.fits} data product issued by the Mikulski Archive for Space Telescopes (MAST) and performs custom background subtraction, flat-field correction, and 1/$f$ noise correction. A high signal-to-noise, normalized reference trace $M_{i,j}$ is then constructed from stacking the out-of-transit corrected 2D images, with $i,j$ the pixels in the $x$ (dispersion) and $y$ (cross-dispersion) directions. Here, $M_{i,j}$ is limited to 30 pixels in the $y$-axis. In the \texttt{SOSSISSE} framework, flux is expressed as a linear combination of $M_{i,j}$ and its spatial derivatives tracing instrumental morphological changes of the trace (Equation A1, \citealt{Lim_2023}), e.g., $x$- and $y$-displacement, rotation, PSF width changes. We utilized two outputs from \texttt{SOSSISSE}: the amplitude term $a(t)$, scaling $M_{i,j}$ at each integration equivalent to the (normalized) stellar flux integrated over all pixels (`white' light curve), and the extracted 2D spectroscopic time series for the spectrophotometric fits.

The \texttt{supreme-SPOON} pipeline works with the \textit{uncal.fits} files available on MAST. We followed Stages 1, 2, and 3 detailed in \cite{Radica_2023} and extracted the flux using a simple box aperture with a width of 30 pixels. The aperture box size was selected to minimize the variance in the out-of-transit white light curve. We expect a negligible level of order contamination (order 2 onto order 1) at wavelengths above 1.6\,$\mu$m for this target \citep{Darveau-Bernier_2022, Radica_2022}. A 1/$f$ noise correction is applied following the method described in \cite{Radica_2023, Benneke_2024}. We, moreover, correct for the presence of the dispersed contaminant star in visit 1 using the method of \citet{Radica_2023}. This correction is done after flux extraction, unlike the \texttt{SOSSISSE} treatment of the contaminant which is applied directly on the detector images (Sect.~\ref{sec:observations}).

\begin{figure}[ht!]
\centering
\minipage{0.5\textwidth}
\includegraphics[width=1\linewidth]{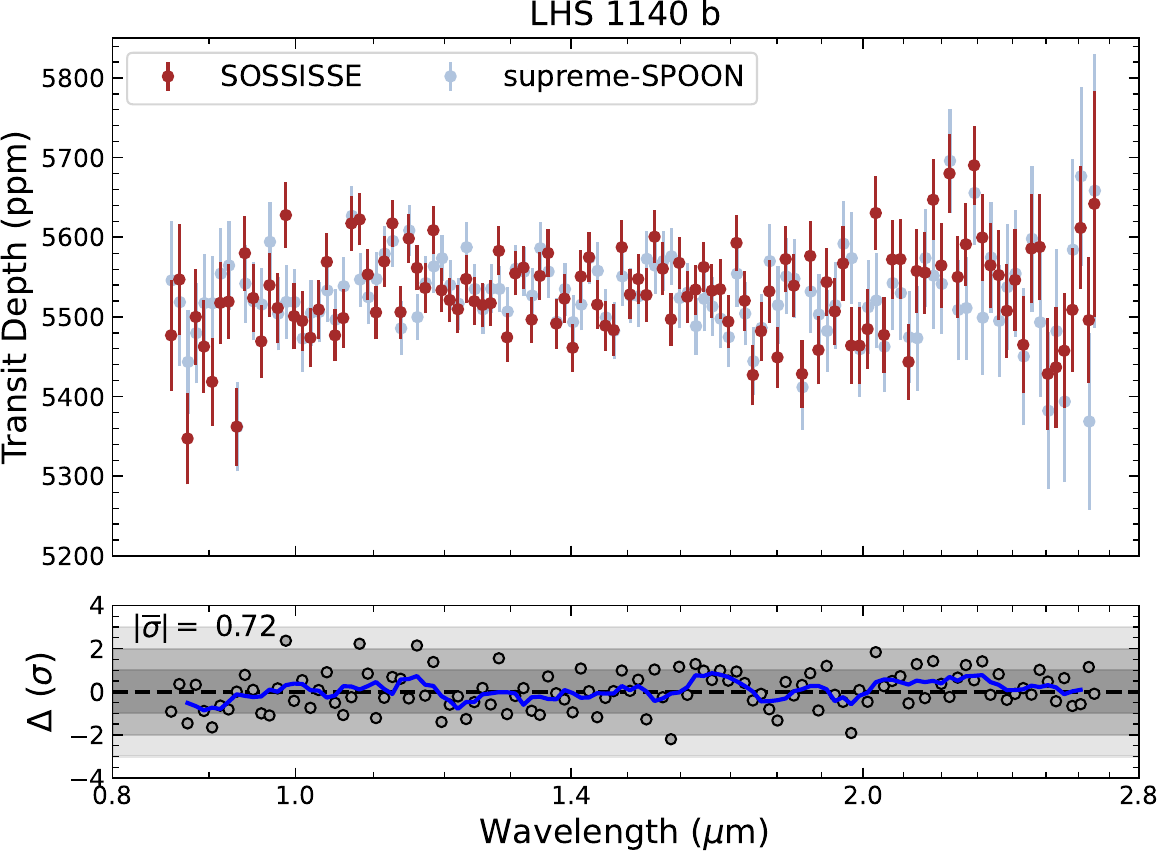}
\endminipage
\hfill
\minipage{0.5\textwidth}
  \includegraphics[width=1\linewidth]{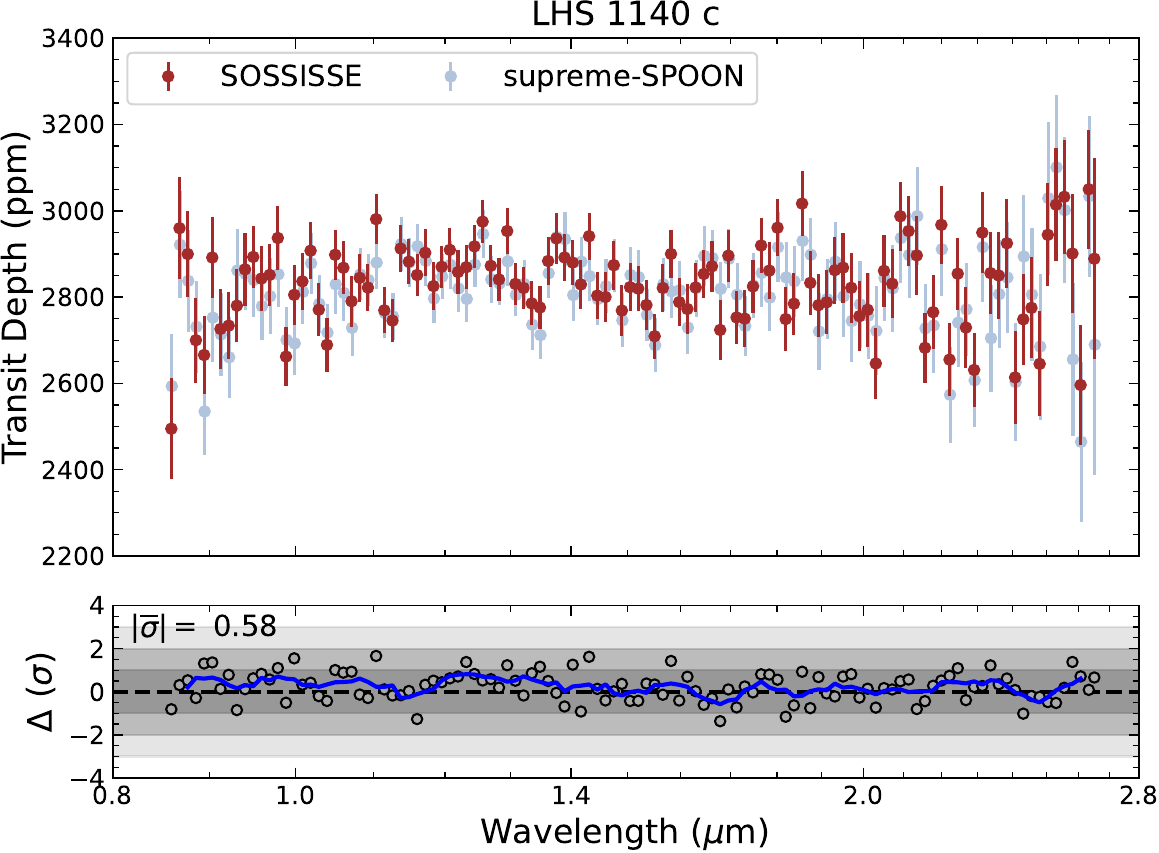}
\endminipage
  \caption{Combined (two visits) transmission spectrum of LHS~1140\,b on the left and (one visit) transmission spectrum of LHS~1140\,c on the right extracted with the \texttt{SOSSISSE} and \texttt{supreme-SPOON} pipelines here showed for the first order of NIRISS/SOSS. The transit light curve fitting is identical between pipelines; any differences arise during the reduction stages. The panels below show the difference in transit depths (\texttt{SOSSISSE} $-$ \texttt{supreme-SPOON}) normalized by the maximum errors between the two pipelines at each spectral bin. A running mean (window = 5) is drawn with a blue curve showing no clear trends in the residuals. The pipelines are in good agreement with an average per-point difference of 0.72 and 0.58\,$\sigma$ for the spectra of LHS~1140\,b and c, respectively.}
  \label{fig:pipeline_diff}
\end{figure}

Figure~\ref{fig:pipeline_diff} shows that the combined transmission spectra at $R \approx 100$ derived from both pipelines are in good agreement with mean (median) absolute deviations of 0.72\,$\sigma$ (0.60\,$\sigma$) for LHS~1140\,b and 0.58\,$\sigma$ (0.52\,$\sigma$) for LHS~1140\,c.

\section{Transmission Spectra} \label{sec:appendix_spectra}
\setcounter{figure}{0}
\renewcommand{\thefigure}{B\arabic{figure}}
\setcounter{table}{0}
\renewcommand{\thetable}{B\arabic{table}}

The final NIRISS SOSS transmission spectra ($R\approx100$) of LHS~1140\,b and c from the combined spectral extraction are given in Table~\ref{table:data}.

\begin{deluxetable}{ccccccc}
\tablecaption{NIRISS SOSS transmission spectra of LHS~1140\,b and c \label{table:data}}
\tablehead{
\colhead{Central wavelength ($\mu$m)} & \colhead{Bandwidth ($\mu$m)} & \colhead{Depth b (ppm)} & \colhead{Error b (ppm)} & \colhead{Depth c (ppm)} & \colhead{Error c (ppm)} & \colhead{Order}
}
\startdata
0.65338 & 0.00327 & 5210.71 & 236.64 & 3331.44 & 362.37 & 2 \\
0.65995 & 0.00330 & 5592.47 & 253.80 & 2247.57 & 397.11 & 2 \\
0.66658 & 0.00333 & 5094.17 & 227.03 & 3275.11 & 391.74 & 2 \\
0.67328 & 0.00337 & 5167.24 & 199.42 & 2676.46 & 374.10 & 2 \\
0.68005 & 0.00340 & 5296.80 & 175.59 & 2650.32 & 358.80 & 2 \\
0.68688 & 0.00343 & 5112.36 & 165.52 & 2645.80 & 321.08 & 2 \\
0.69379 & 0.00347 & 5322.05 & 141.06 & 2833.85 & 259.10 & 2 \\
\ldots & \ldots & \ldots & \ldots & \ldots & \ldots & \ldots \\
\enddata
\tablecomments{Table~\ref{table:data} is available in its entirety in machine-readable format.}
\end{deluxetable}

\section{Preliminary Analysis of the Transmission Data of LHS~1140\,{\footnotesize{c}}}  \label{sec:appendix_lhs1140c}

\setcounter{figure}{0}
\renewcommand{\thefigure}{C\arabic{figure}}
\setcounter{table}{0}
\renewcommand{\thetable}{C\arabic{table}}

\begin{figure}[b!]
\centering
\includegraphics[width=1\linewidth]{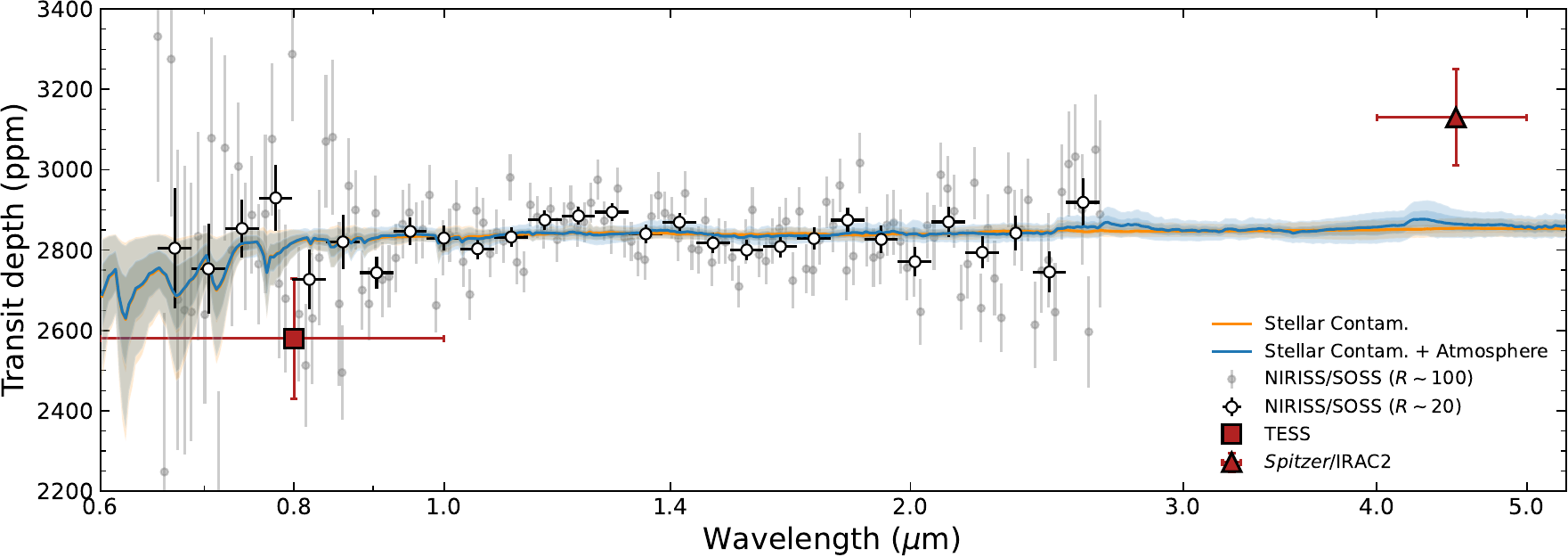}
  \caption{Transmission spectrum of LHS~1140\,c from NIRISS/SOSS at a $R\sim20$ (white points) and at a higher $R\sim100$ (gray points). The best-fit models from \texttt{POSEIDON} with 1-2$\sigma$ confidence envelopes are shown for a stellar contamination model (orange) and a joint stellar contamination and planetary atmosphere model (blue). The TESS and \textit{Spitzer} photometric measurements \citep{Cadieux_2024} are shown in red. We have included the \textit{Spitzer} point in the retrieval analysis, but its exclusion does not change the results. The NIRISS/SOSS observations is flat with marginal evidence of unocculted faculae consistent with the simultaneous transit of LHS~1140\,b (Fig.~\ref{fig:transmission_spectrum}). The spectrum covering 0.65--2.7\,$\mu$m has an average transit depth that falls between those measured by TESS and \textit{Spitzer}, resolving the 4\,$\sigma$ discrepancy in radius measurements between the two instruments noted by \cite{Cadieux_2024}.}
  \label{fig:transmission_spectrum_lhs1140c}
\end{figure}

The first transmission spectrum of the warm super-Earth LHS~1140\,c is presented in Figure~\ref{fig:transmission_spectrum_lhs1140c}. We present here a first exploratory analysis of LHS~1140\,c's NIRISS/SOSS spectrum by repeating the retrieval exercise outlined in Section~\ref{sec:retrieval} for the Flat line, TLS-only, and TLS + Multi-gas (Haze) models. All priors remain unchanged except for $R_{\rm p,ref}$ and $M_{\rm p}$ changed to $\mathcal{U}\left(1, 1.5\right)$ and $\mathcal{N}\left( 1.91, 0.06^2\right)$ based on \cite{Cadieux_2024}, and $T$ now allowed to go to higher temperatures (100--600\,K).

A flat solution is favored for LHS~1140\,c (2.1\,$\sigma$) due to the lower signal-to-noise ratio spectrum derived from a single transit entirely confined into a transit of LHS~1140\,b (double transit, Fig.~\ref{fig:spectral_fit}). The Flat line model yields a $\ln \mathcal{Z} = 1090.49$, while the TLS-only and TLS + Multi-gas (Haze) retrieval runs end up with marginally lower Bayesian evidences of $\ln \mathcal{Z} = 1089.34$ and 1088.83, respectively.

The stellar contamination only model yields consistent unocculted heterogeneity properties compared to LHS~1140\,b: $f_{\rm fac} = 0.12^{+0.18}_{-0.07}$ and $T_{\rm fac} = 3293^{+205}_{-111}$\,K, with $f_{\rm spot}$ consistent with 0. The TLS effect associated with LHS~1140\,c is smaller by definition given the surface ratio of the planets $(R_{\rm p, c} / R_{\rm p, b})^2$ of about one half.

Our joint stellar contamination and planetary atmosphere retrieval (TLS + Multi-gas model) can inform on possible atmospheres on LHS~1140\,c. The posterior distributions for this run are presented in Appendix~\ref{sec:appendix_supplementary}. We do not detect any molecular feature or a slope from hazes (haze parameters are unconstrained). The most likely composition favored by the data is a pure N$_2$ envelope at any $T$. Two other solutions exist that are approximately equally likely: a H$_2$O-rich atmosphere at $T\approx200$\,K or a CO$_2$-rich atmosphere at $T \lesssim 500$\,K. The combination of a high mean molecular weight $\mu$ and/or low $T$ for such atmospheres is sufficient to flatten the spectrum below the sensitivity of our NIRISS spectrum. The data is incompatible with clear CH$_4$-rich and H$_2$-rich atmospheres with 2$\sigma$ upper limits of $\log \mathrm{CH_4} < -2.25$ and $\log \mathrm{H_2} < -0.94$ ($<$ 11\%), respectively.

\section{Supplementary Material: Atmospheric Inference Analysis} \label{sec:appendix_supplementary}

\setcounter{figure}{0}
\renewcommand{\thefigure}{D\arabic{figure}}
\setcounter{table}{0}
\renewcommand{\thetable}{D\arabic{table}}

\subsection{Evidence for a N$_2$-Dominated Atmosphere on LHS~1140\,{\footnotesize{b}}}  \label{sec:appendix_lhs1140b_n2_evidence}

Here, we elucidate why our retrievals favor an N$_2$-dominated atmosphere. The observed transmission spectrum is given by $\Delta_{\lambda, \, \rm{obs}} = \epsilon_{\lambda, \, \rm{star}} \, \Delta_{\lambda, \, \rm{atm}}$, where $\epsilon_{\lambda, \, \rm{star}}$ is the `stellar contamination factor' \citep[e.g., Equation 3 in][]{Fournier-Tondreau_2024} and $\Delta_{\lambda, \, \rm{atm}}$ is the `regular' transmission spectrum of the planetary atmosphere. Given that unocculted faculae are the dominant wavelength-dependent effect sculpting LHS~1140\,b's transmission spectrum, we first corrected our NIRISS/SOSS data by dividing out the best-fitting unocculted stellar contamination model (binned to the resolution of the data). This allows one to compare model atmosphere-transmission-only spectra, $\Delta_{\lambda, \, \rm{atm}}$, to an equivalent form of the data.

\begin{figure}[b!]
\centering
\includegraphics[width=0.8\linewidth]{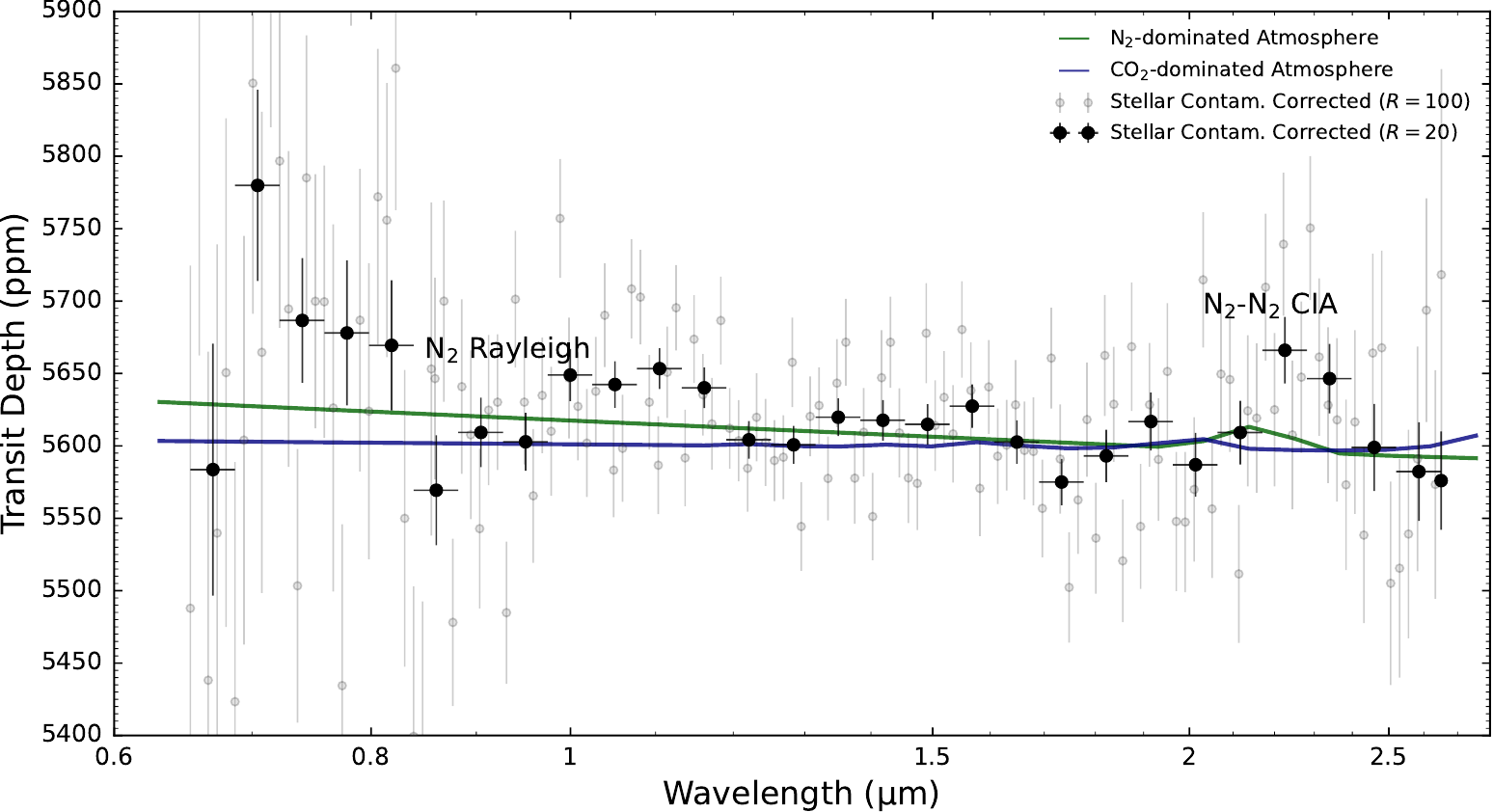}
  \caption{Tentative evidence of an N$_2$-dominated atmosphere on LHS~1140\,b. The stellar contamination corrected LHS~1140\,b NIRISS/SOSS transmission spectrum (black points for $R \sim 20$, gray points for $R \sim 100$) is compared to the best-fitting model transmission spectra for a 100\% N$_2$ atmosphere (green) and a 100\% CO$_2$ atmosphere (blue), plotted at $R = 20$. The N$_2$ model has a best-fitting $T = 399$\,K, while the CO$_2$ model has $T = 116$\,K. The N$_2$ model provides a better fit (2.3\,$\sigma$) due to the presence of a Rayleigh scattering slope and a weak N$_2$-N$_2$ collision-induced absorption feature.}
  \label{fig:TLS_Corrected_Transmission_Spectrum}
\end{figure}

Figure~\ref{fig:TLS_Corrected_Transmission_Spectrum} compares our stellar contamination-corrected transmission spectrum of LHS~1140\,b to our best-fitting 100\% N$_2$ and 100\% CO$_2$ atmosphere models. The CO$_2$ model is essentially flat, consistent with a non-detection of CO$_2$ bands. The N$_2$ model, however, exhibits a clear Rayleigh slope in both the first and second orders of the NIRISS/SOSS data (alongside a weaker N$_2$-N$_2$ collision-induced absorption feature near 2.2\,$\mu$m). Both models allow for the possibility of power-law hazes, but N$_2$ Rayleigh scattering alone provides the necessary slope without any additional scattering opacity. We stress that the residual slope attributed here to N$_2$ relies on the assumption that the stellar contamination is well-described by our interpolated PHOENIX models for faculae. The best TLS model leaves a significant residual near 2.3\,$\mu$m which is coincident with the CO bandhead, a well-known temperature- and gravity-sensitive spectral feature in M dwarf spectra. Given inherent uncertainties associated with stellar atmosphere models \citep[e.g.,][]{Lim_2023, Jahandar2024}, one cannot rule out the possibility that the measured slope is an artifact of imperfect stellar contamination correction. The evidence for N$_2$ is only tentative at this stage (2.3\,$\sigma$) and will need to be confirmed with future observations.

\subsection{Atmospheric and Stellar Retrieval Results}

\begin{figure}[ht!]
\centering
\includegraphics[width=1\linewidth]{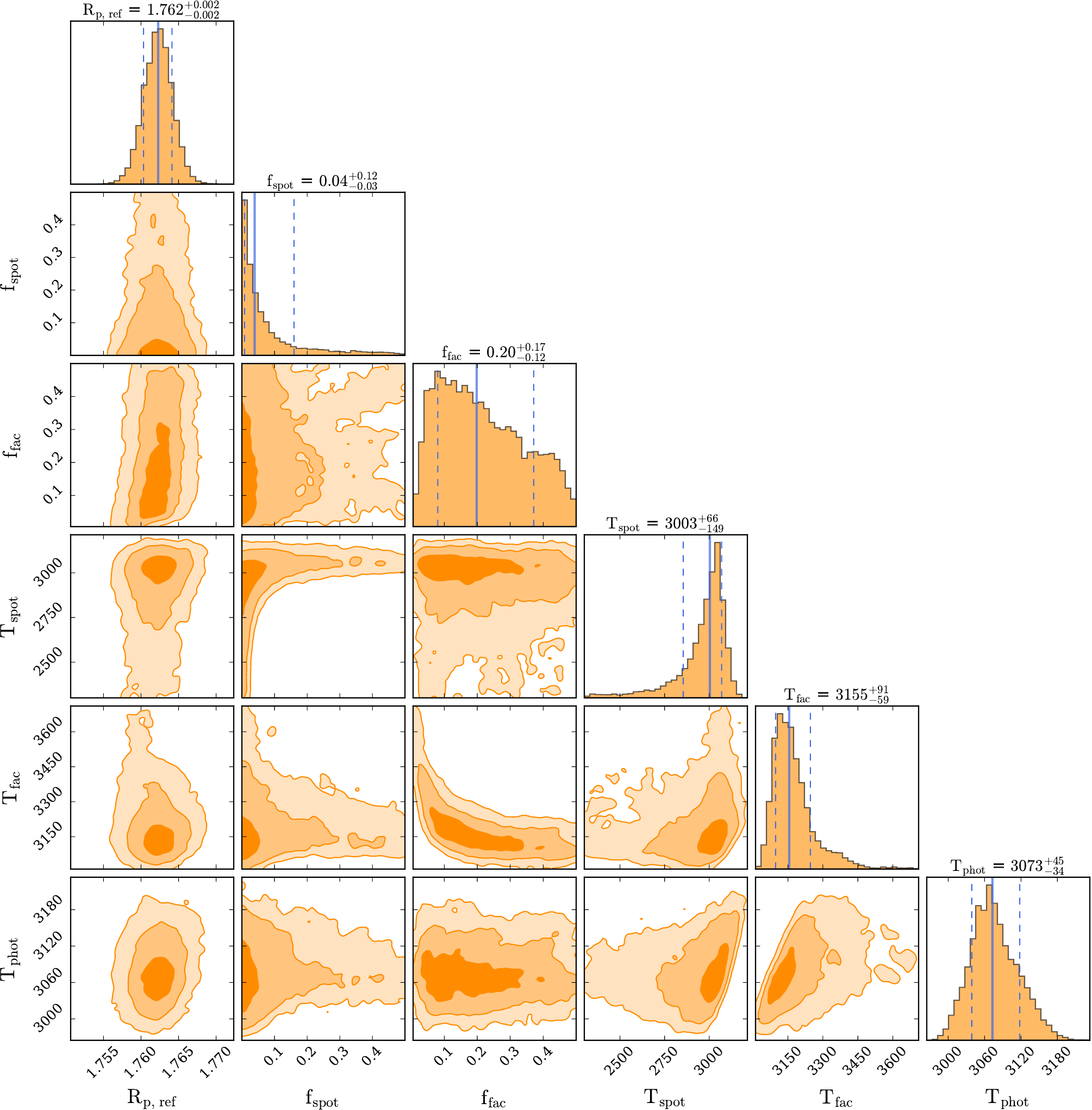}
  \caption{Posterior distribution for the stellar contamination only (`TLS-only') \texttt{POSEIDON} retrieval of LHS~1140\,b's combined NIRISS/SOSS transmission spectrum.}
  \label{fig:corner_tls}
\end{figure}

\begin{figure}[ht!]
\centering
\includegraphics[width=1\linewidth]{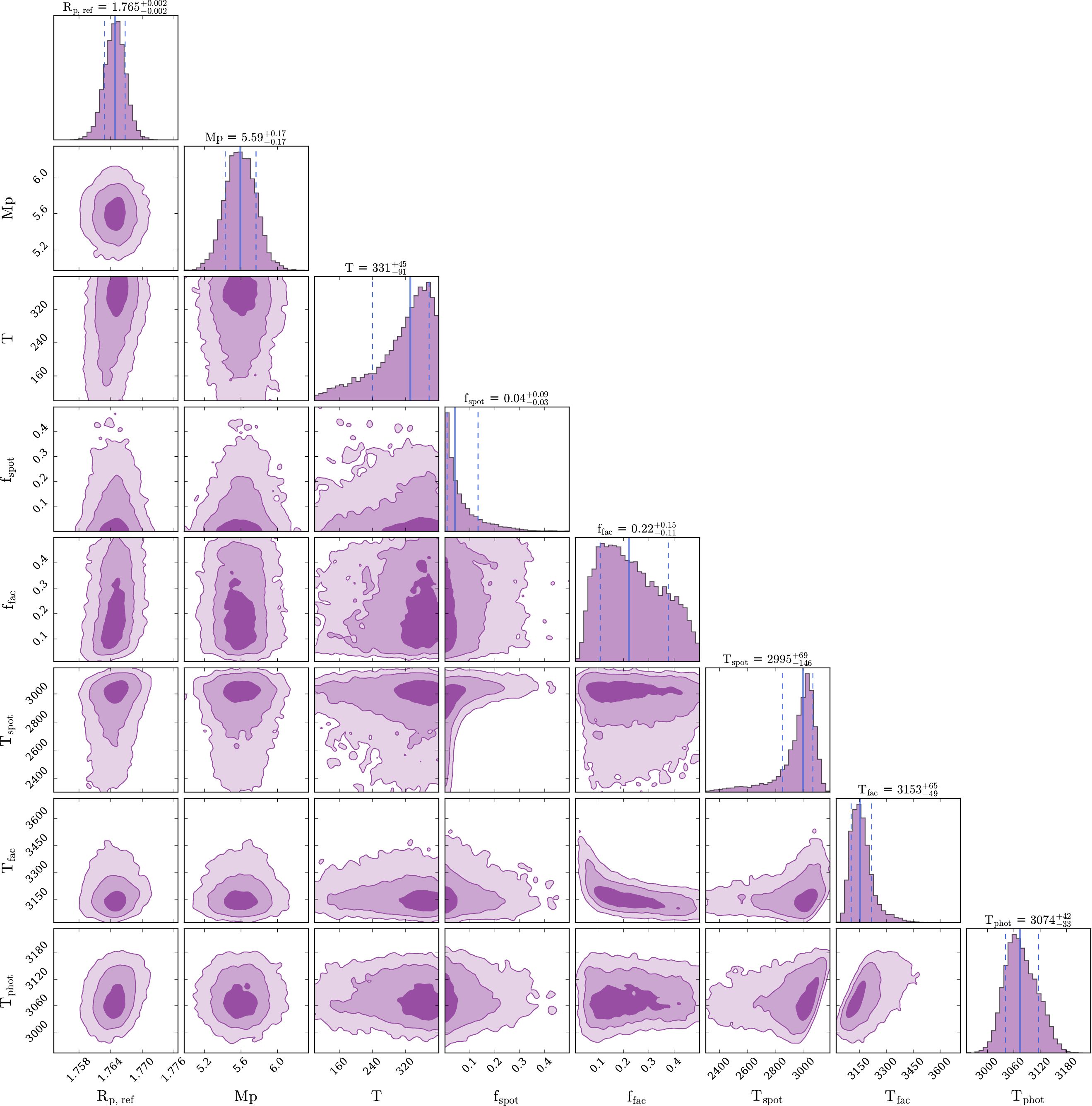}
  \caption{Posterior distribution for the joint N$_2$ atmosphere and stellar contamination retrieval of LHS~1140\,b's combined NIRISS/SOSS transmission spectrum. Note that N$_2$ does not show in the corner plot as a free parameter because the N$_2$ abundance is fixed to 100\% for this model.}
  \label{fig:corner_tls+n2}
\end{figure}

\begin{figure}[ht!]
\centering
\includegraphics[width=1\linewidth]{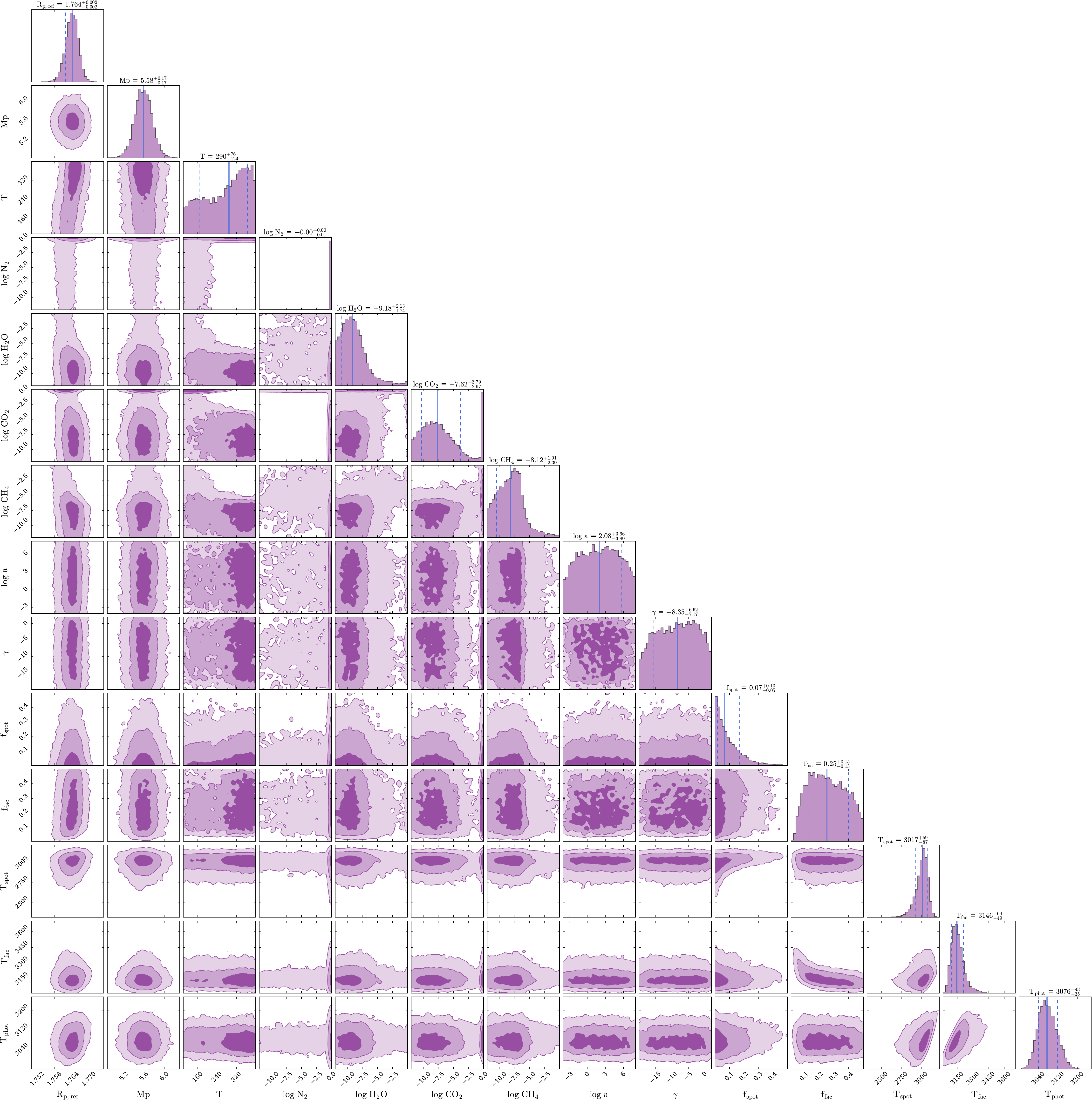}
  \caption{Posterior distribution for the joint multi-gas atmosphere with hazes and stellar contamination retrieval of LHS~1140\,b's combined NIRISS/SOSS transmission spectrum.}
  \label{fig:corner_tls+atmosphere}
\end{figure}

\begin{figure}[ht!]
\centering
\includegraphics[width=1\linewidth]{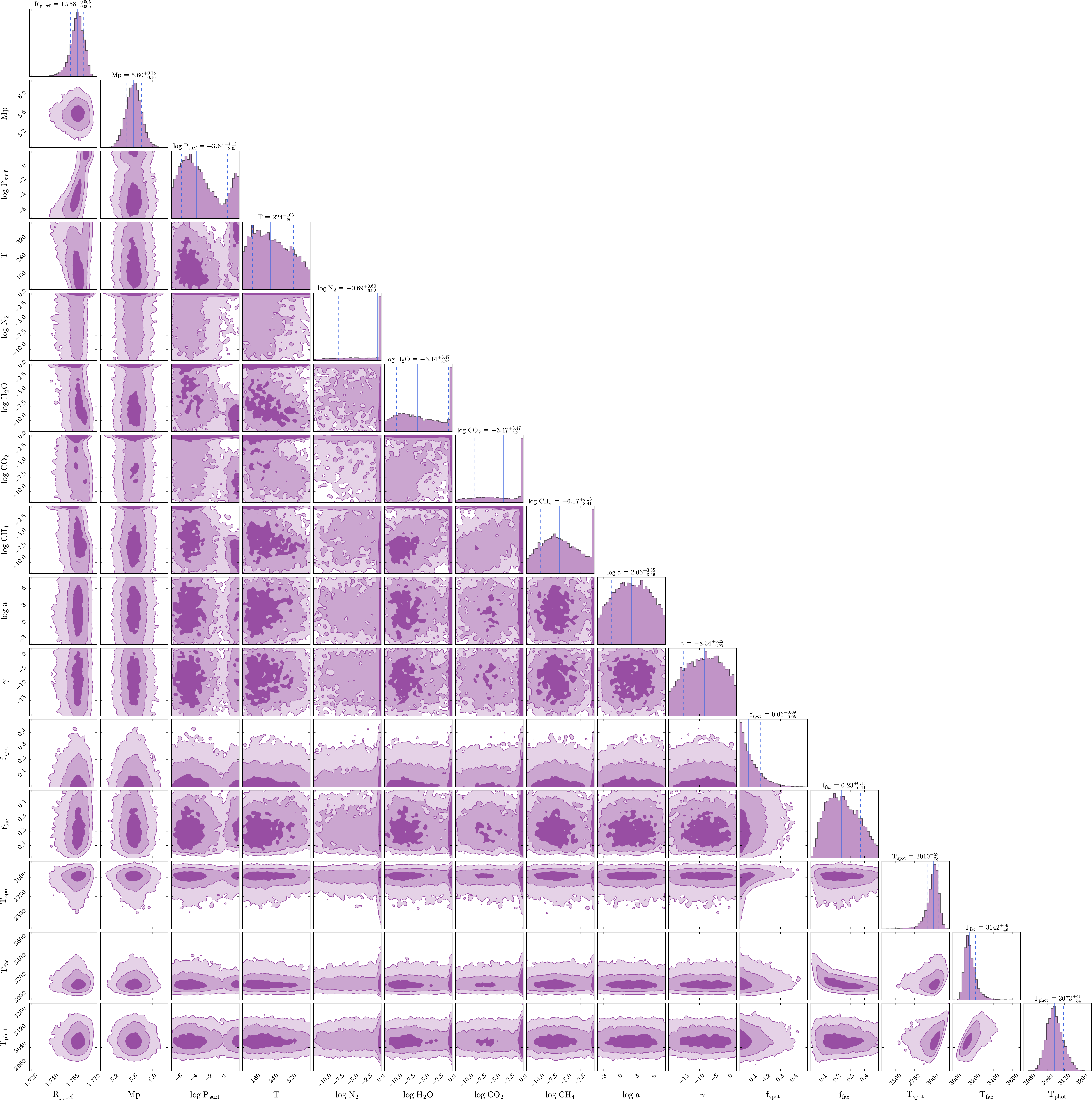}
  \caption{Posterior distribution for the joint multi-gas atmosphere with hazes, a cloud/surface, and stellar contamination retrieval of LHS~1140\,b's combined NIRISS/SOSS transmission spectrum.}
  \label{fig:corner_tls+atmosphere_cloudy}
\end{figure}

\begin{figure}[ht!]
\centering
\includegraphics[width=1\linewidth]{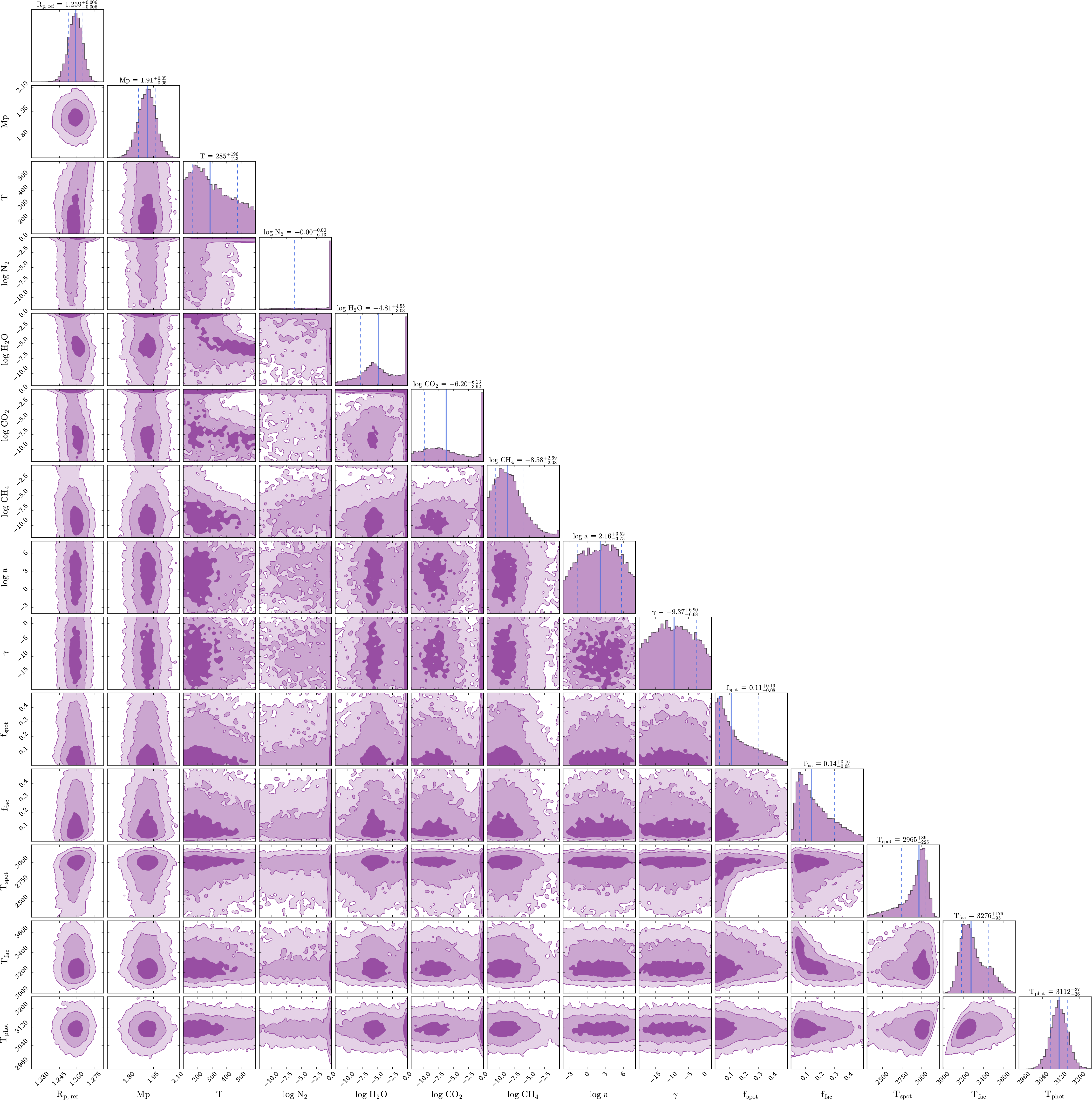}
  \caption{Posterior distribution for the joint multi-gas atmosphere with hazes and stellar contamination of LHS~1140\,c's NIRISS/SOSS transmission spectrum.}
  \label{fig:corner_lhs1140c}
\end{figure}

The priors, posterior parameter constraints, and key statistics from our retrieval analysis of LHS~1140\,b are presented in Table~\ref{table:retrieval_results}. The full posterior distributions are shown for the TLS-only model in Figure~\ref{fig:corner_tls}, TLS + N$_2$ (Clear) in Figure~\ref{fig:corner_tls+n2}, TLS + Multi-gas (Haze) in Figure~\ref{fig:corner_tls+atmosphere}, and TLS + Multi-gas (Haze + Cloud) in Figure~\ref{fig:corner_tls+atmosphere_cloudy}. Finally, the posterior distributions for the TLS + Multi-gas (Haze) retrieval of LHS~1140\,c are presented in Figure~\ref{fig:corner_lhs1140c}.

\begin{table}[b!]
\end{table}
\begin{deluxetable}{lcccccc}
\tablecaption{Retrieval analysis summary for LHS~1140\,b's combined NIRISS/SOSS transmission spectrum}
\tablehead{\colhead{Parameter} & \colhead{Prior} & \colhead{Flat line} & \colhead{TLS only} & \colhead{TLS + N$_2$} & \colhead{TLS + Multi-gas} & \colhead{TLS + Multi-gas}\\[-0.3cm]
& & & & \colhead{(Clear)} & \colhead{(Haze)} & \colhead{(Haze + Cloud)}}
\startdata \noalign{\vskip 5pt}
\multicolumn{2}{c}{\textit{Stellar Contamination}} & & & & & \\ \noalign{\vskip 5pt}
$T_{\rm spot}$ (K) & $\mathcal{U}\left(2300, T_{\rm eff} + 3\,\sigma_{T_{\rm eff}}\right)$ & -- & 3003$^{+66}_{-149}$ & 2995$^{+69}_{-146}$ & 3017$^{+60}_{-87}$ & 3010$^{+59}_{-88}$ \\
$T_{\rm phot}$ (K) & $\mathcal{N}\left(T_{\rm eff}, \sigma_{T_{\rm eff}}^2 \right)$ & -- & 3073$^{+45}_{-34}$ & 3074$^{+42}_{-33}$ & 3076$^{+43}_{-35}$ & 3073$^{+41}_{-34}$ \\
$T_{\rm fac}$ (K) & $\mathcal{U}\left(T_{\rm eff} - 3\,\sigma_{T_{\rm eff}}, 1.2\,T_{\rm eff}\right)$ & -- & 3155$^{+91}_{-59}$ & 3153$^{+65}_{-49}$ & 3146$^{+64}_{-49}$ & 3142$^{+66}_{-46}$ \\
$f_{\rm spot}$ & $\mathcal{U}\left(0, 0.5\right)$ & -- & 0.04$^{+0.12}_{-0.03}$ & 0.04$^{+0.09}_{-0.03}$ & 0.07$^{+0.10}_{-0.05}$ & 0.06$^{+0.09}_{-0.05}$ \\
$f_{\rm fac}$ & $\mathcal{U}\left(0, 0.5\right)$ & -- & 0.20$^{+0.17}_{-0.12}$ & 0.22$^{+0.15}_{-0.11}$ & 0.25$^{+0.15}_{-0.13}$ & 0.23$^{+0.14}_{-0.11}$  \\ \noalign{\vskip 5pt}
\multicolumn{2}{c}{\textit{Atmospheric Properties}}  & & & & & \\ \noalign{\vskip 5pt}
$R_{\rm p, ref}$ (R$_{\oplus}$) & $\mathcal{U}\left(1.5,  2.0\right)$ & 1.7513$\pm$0.0006 & 1.762$\pm$0.002 & 1.765$\pm$0.002 & 1.764$\pm$0.002 & 1.758$\pm$0.005 \\
$M_{\rm}$ (M$_{\oplus}$) & $\mathcal{N}\left(5.60, 0.19^2 \right)$ & -- & -- & 5.59$\pm$0.17 & 5.48$\pm$0.17 & 5.60$\pm$0.16 \\
$T$ (K) & $\mathcal{U}\left(100,  400\right)$ & -- & -- & 331$^{+45}_{-91}$ & 290$^{+76}_{-124}$ & 224$^{+103}_{-80}$ \\
$\log a_{\rm haze}$ & $\mathcal{U}\left(-4,  8\right)$ & -- & -- & -- & $2.08^{+3.65}_{-3.80}$ & $2.06^{+3.55}_{-3.56}$ \\
$\gamma_{\rm haze}$ & $\mathcal{U}\left(-20,  2\right)$ & -- & -- & -- & $-8.35^{+6.52}_{-7.17}$ & $-8.34^{+6.32}_{-6.77}$ \\
$\log (P_{\rm surf}/$bar) & $\mathcal{U}\left(-7,  2\right)$ & -- & -- & -- & -- & $-3.64^{+4.12}_{-2.05}$ \\
$\log$ {X$_{\textrm{\tiny H$_2$}}$} & (derived) & -- & -- & -- & $< -0.95$ & $< -0.56$ \\
$\log$ {X$_{\textrm{\tiny N$_2$}}$} & $\mathcal{CLR}\left(-12,  0\right)$ & -- & -- & 0 (fixed) & -0.00$^{+0.00}_{-0.01}$ & -0.69$^{+0.69}_{-6.92}$ \\
$\log$ {X$_{\textrm{\tiny CO$_2$}}$} & $\mathcal{CLR}\left(-12,  0\right)$ & -- & -- & --  & -7.62$^{+3.79}_{-2.67}$ $^{\dagger}$ & -3.47$^{+3.47}_{-5.25}$ \\
$\log$ {X$_{\textrm{\tiny H$_2$O}}$} & $\mathcal{CLR}\left(-12,  0\right)$ & -- & -- & -- & $<-2.94$ & -6.14$^{+5.47}_{-3.74}$ \\
$\log$ {X$_{\textrm{\tiny CH$_4$}}$} & $\mathcal{CLR}\left(-12,  0\right)$ & -- & -- & -- & $<-2.78$ & -6.17$^{+4.16}_{-3.41}$  \\ \noalign{\vskip 5pt}
\hline \noalign{\vskip 5pt}
\multicolumn{3}{c}{\textit{Model Statistics}} & & & & \\ \noalign{\vskip 5pt}
 & $N_{\rm params}$ & 1 & 6 & 8 & 14 & 15 \\
 & $\chi^2$ (d.o.f.) & 277 (141) & 235 (136) & 228 (134) & 228 (128) & 228 (127) \\
 & $\chi^2_{\nu}$ & 1.96 & 1.73 & 1.70 & 1.78 & 1.80 \\
 & $\ln \mathcal{Z}$ & 1127.24 & 1142.15 & 1144.65 & 1141.79 & 1141.60 \\
\enddata
\tablecomments{TLS = `Transit Light Source Effect' (contamination from unocculted stellar active regions, \citealt{Rackham_2018}). `Multi-gas' retrievals include N$_2$, H$_2$, CO$_2$, H$_2$O, and CH$_4$. $\mathcal{CLR}$ refers to the centered-log ratio prior \citep[e.g.,][]{Benneke_2012}. H$_2$ also follows a $\mathcal{CLR}$ prior, even though it is not a free parameter (it is derived from the summation to unity condition), since this is defining feature of the centered-log ratio. The \emph{a priori} known stellar properties are $T_{\rm eff} = 3096$ and $\sigma_{T_{\rm eff}} = 48$\,K \citep{Cadieux_2024}. $\dagger$ CO$_2$ has a biomodal distribution, with either (i) a non-detection for a pure N$_2$ atmosphere or (ii) 100\% CO$_2$ to render the spectrum flat (the latter solution is statistically disfavored). Upper bounds correspond to 95\% confidence (2$\sigma$) limits. Figure~\ref{fig:transmission_spectrum} shows results for the `TLS only' (orange histograms) and the `TLS + Multi-gas (Haze)' (blue histograms) models summarized in this Table.}
\label{table:retrieval_results}
\end{deluxetable}

\clearpage

\section{Global Climate Model Simulations and Synthetic Observables} \label{sec:appendix_gcm}

\setcounter{figure}{0}
\renewcommand{\thefigure}{E\arabic{figure}}
\setcounter{table}{0}
\renewcommand{\thetable}{E\arabic{table}}

We performed a series of 3-D GCM (Global Climate Model) simulations designed to explore a large variety of plausible atmospheric compositions for LHS~1140\,b. These simulations include (i) thick mini-Neptune atmospheres (80~bar H$_2$-rich) with compositions of 100$\times$, 300$\times$ and 1000$\times$solar metallicity, described in the main text, (ii) compact secondary N$_2$ and CO$_2$-rich atmospheres, described in \cite{Cadieux_2024}, and (iii) sensitivity experiments of H$_2$-rich atmospheres on top of a dry surface or a global surface ocean a.k.a.\ `hycean'-type planet \citep{Madhusudhan_2021}.

\begin{figure}[hb!]
\centering
\includegraphics[width=0.9\linewidth]{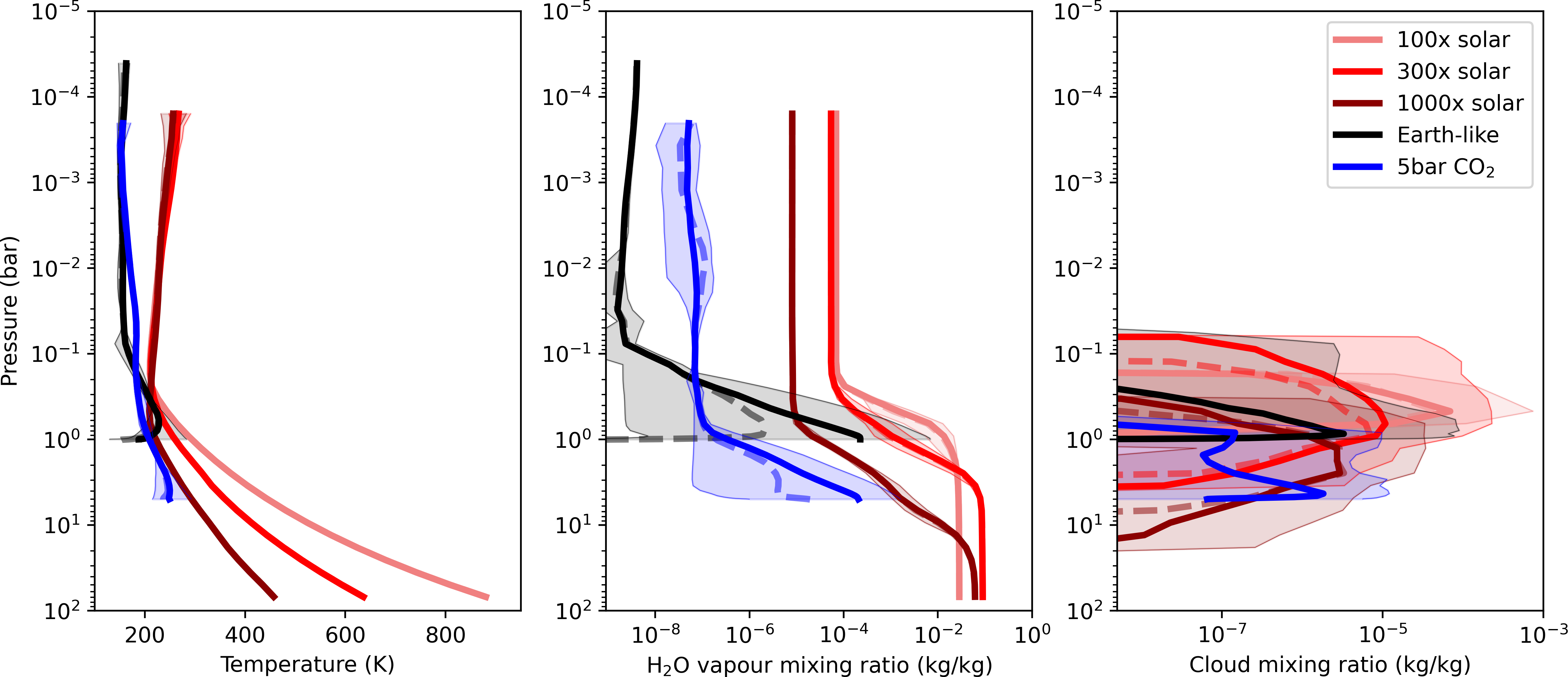}
  \caption{Results of Global Climate Model (GCM) simulations of LHS~1140\,b assuming mini-Neptune thick H$_2$ atmosphere (100$\times$, 300$\times$ and 1000$\times$solar metallicity composition), an Earth-like atmosphere (1\,bar N$_2$, 400\,ppm CO$_2$), and a thick CO$_2$-dominated atmosphere (5\,bar of CO$_2$). Panels show vertical profiles of the atmospheric temperatures (left), water vapor mixing ratio (middle), and water cloud mixing ratios (right). The thick solid lines indicate the global mean vertical profiles, and the dotted lines indicate the terminator vertical profiles (impacting transit spectra).}
  \label{fig:PTprofiles_GCM}
\end{figure}

Figure~\ref{fig:PTprofiles_GCM} shows the temperature, water vapor mixing ratio and cloud profiles for the mini-Neptune cases, as well as for the Earth-like and CO$_2$-dominated atmospheres for comparison. In all these simulations, cloud decks are located below atmospheric pressures of $\sim$0.1\,bar, near the top of the tropospheric cold trap. Clouds therefore have a very limited impact on the transit spectra (see Fig.~\ref{fig:forward_models}) for all the simulations explored in this work. Previous intercomparison works \citep{Sergeev_2022,Fauchez_2022} have shown that cloud deck altitude can vary slightly from one model to another, but here cloud deck altitude would have to rise by at least two orders of magnitude to alter our conclusions. The presence of high-altitude hazes --- which is not simulated here --- could flatten the transit spectrum even in the case of an atmosphere rich in H$_2$, but in this scenario we would expect to see a haze slope in the shortwave part of spectrum \citep{Sing_2011}.

Some simulated cases show a slight variability in the amplitude of transit spectra due to cloud variability (in time and location), as already shown in \citealt{Charnay_2021} in the case of K2-18\,b. However, even at model timesteps where the terminator region is the cloudiest, the amplitude of the transit spectrum features for the mini-Neptune cases is far greater than the scatter of NIRISS observations, so these compositions can be formally rejected.

We ran additional sensitivity simulations in which we changed the boundary conditions for the mini-Neptune simulations. First, we added a dry surface at 10\,bar atmospheric pressure, but in this case the synthetic spectra (not shown) are very similar to the mini-Neptune cases (see Fig.~\ref{fig:forward_models}).  Second, we added a surface entirely covered by an ocean also at 10\,bar atmospheric pressure. This aquaplanet endowed with a thick H$_2$-dominated atmosphere, also known as hycean planet, has a quite different behavior. In fact, the simulation enters runaway greenhouse, which forces the ocean to evaporate and the surface temperatures to rise. This result is compatible with previous calculations of the runaway greenhouse limit for hycean planets \citep{Innes_2023}. As the simulation progresses, the amount of water vapor in the atmosphere increases at all altitude layers. The altitude of the water cloud deck also increases, similar to what has been shown already for N$_2$-dominated atmospheres entering the runaway greenhouse \citep{Chaverot_2023}. For all the hycean cases we have simulated, the amplitude of the spectral atmospheric features is far greater than the dispersion of the NIRISS/SOSS spectrum, even accounting for the effect of clouds. However, these runaway greenhouse simulations have by definition not reached their final equilibrium state, with clouds potentially forming even higher in the atmosphere. Moreover, it is not known whether the ocean at the base of the H$_2$-H$_2$O-dominated atmosphere can eventually stabilize at higher temperatures, which would also likely affect the water vapor and cloud structures. Last but not least, we have not accounted for convection inhibition, which would likely make this type of atmosphere much hotter in the lowest atmospheric layers, and even less likely for an ocean to exist \citep{Innes_2023,Leconte_2024}. Performing fully converged 3-D GCM simulations of hycean planets, accounting for moist convection, clouds, and convection inhibition together at the same time, is a challenging yet necessary step \citep{Leconte_2024} to simulate accurately the evolution and stability of LHS~1140\,b as an hycean planet.

\end{document}